\documentclass{aa}  

\usepackage{graphicx}
\usepackage{arydshln}
\usepackage{txfonts}
\usepackage{placeins}
\begin{document}

   \title{A Momentum-Regulated Model\\For Star Formation Efficiency in Giant Molecular Clouds}

   \author{E. Bertram}

   \institute{Hochschule Fresenius Heidelberg, Sickingenstra\ss{}e 63-65, 69126 Heidelberg, Germany\\ \email{mail@erik-bertram.de}}

   \date{Accepted July 7, 2026}

\abstract{We present a minimal analytic framework to investigate the star formation efficiency per free-fall time, $\epsilon_{\rm ff}$, in giant molecular clouds (GMCs), focusing on the origin of the observed clustering around $\epsilon_{\rm ff} \sim 0.01$. We model the time evolution of the turbulent velocity dispersion through a momentum balance between stellar feedback and turbulent dissipation, and show that this generically leads to a stable low-efficiency equilibrium with only weak dependence on global cloud properties. We extend the framework by including a phenomenological contribution from gravity-driven turbulence and find that both feedback- and gravity-driven motions converge to similar equilibrium states under typical GMC conditions. The efficiency can be expressed as the ratio between a gravitational velocity scale and an effective feedback velocity scale, providing a physically transparent interpretation of self-regulated star formation. The model provides a simple, physically motivated interpretation of observed gas--star formation scaling relations, including a Schmidt-like scaling at cloud scales and a Kennicutt-like scaling when averaged over cloud populations. Comparison with observed GMC properties shows agreement within a factor of a few and highlights the weak sensitivity of $\epsilon_{\rm ff}$ to cloud parameters. Despite its simplicity, the framework captures the leading-order interplay between turbulence, gravity, and feedback, and provides a physically transparent explanation for the origin and robustness of low star formation efficiencies in GMCs.}


\maketitle
\nolinenumbers

%

\section{Introduction}

Star formation in galaxies is observed to proceed remarkably inefficiently. If self-gravitating molecular gas were to collapse
on a global free-fall time without additional support, the implied star formation rates would exceed observed values by more than an order of magnitude \citep{Zuckerman1974, McKee1997, Robitaille2010}. Observations of giant molecular clouds (GMCs) in the Milky Way and nearby galaxies instead indicate that only a small fraction of the available gas is converted into stars per free-fall time, typically $\epsilon_{\rm ff} \sim 0.5$--$2\%$ \citep{Bonnell2011, Krumholz2012, Evans2014, Heyer2016, Lee2016, Ochsendorf2017, Barnes2017, Chevance2019}. Thus, understanding the origin of this remarkably low efficiency remains one of the central problems in current star formation theory.

A broad class of theoretical models explains low star formation efficiencies (SFE) through the statistical properties of supersonic turbulence. In these turbulence-regulated models, only the highest-density fraction of the gas distribution is able to collapse gravitationally, leading to low average efficiencies \citep{Krumholz2005, PadoanNordlund2011, HennebelleChabrier2011, Padoan2012, Federrath2012, Federrath2013, Faucher2013, Federrath2015, Bertram2016}. Magnetic fields, intermittent forcing, and variations in the turbulent Mach number further modify the predicted collapse fraction and can significantly affect the resulting star formation rates \citep{Federrath2012, Federrath2015}. These models successfully reproduce many statistical properties of turbulent star-forming gas and have provided important insight into the density structure of the interstellar medium (ISM).

A complementary theoretical framework emphasizes the role of stellar feedback in regulating turbulence and maintaining
dynamical equilibrium within the ISM. In particular, \citet{Ostriker2011} demonstrated that momentum injection from supernovae can balance vertical gravity and turbulent dissipation in galactic discs, leading to self-regulated star formation laws. \citet{Faucher2013} extended these ideas by connecting cloud-scale star formation to galaxy-scale gas regulation and deriving Kennicutt--Schmidt scaling relations from feedback-regulated equilibrium arguments. More recently, \citet{Ostriker2022} developed the pressure-regulated, feedback-modulated (PRFM) framework, which provides a unified description of self-regulated star formation calibrated against numerical simulations. Together, these studies established stellar feedback as a key mechanism regulating turbulence and star formation in the multiphase ISM.

Numerical simulations have been instrumental in exploring star formation in turbulent, magnetized, and feedback-regulated
environments \citep{Hopkins2014, Semenov2018, Federrath2015, Grudic2018, Kim2021}. Multi-physics simulations including supernovae, stellar winds, radiation pressure, and magnetic fields consistently find that both the instantaneous and integrated star formation efficiency depend sensitively on the balance between gravity and feedback, as well as on gas surface density and cloud structure \citep{Federrath2015, Grudic2018}. At the same time, simulations also indicate that the efficiency is not strictly universal and may vary systematically with cloud environment, evolutionary stage, or local ISM conditions \citep{Semenov2015}. These studies reinforce the importance of dynamical self-regulation while also highlighting the need for simplified analytic models that clarify the dominant physical dependencies.

At the same time, several studies have argued that turbulence may also be driven directly by gravitational processes. In particular, \citet{Krumholz2018} proposed a unified model in which turbulence in galactic discs is sustained jointly by stellar feedback and gravitational energy released by radial transport and gravitational instability. Extensions of this framework further incorporated additional driving channels such as accretion \citep{Klessen2010} and cosmological inflows \citep{Ginzburg2022, Forbes2023}. Numerical simulations likewise demonstrate that gravitational collapse itself can generate turbulent motions through converging flows and hierarchical contraction within molecular clouds \citep{Robertson2012, Murray2015}. These studies suggest that both stellar feedback and gravity may contribute simultaneously to the turbulent energy budget of star-forming systems.

Observational studies likewise suggest that the limited variation of $\epsilon_{\rm ff}$ may only be approximate. Resolved surveys of molecular gas in nearby galaxies find typical efficiencies of order $\sim 1\%$, but with significant cloud-to-cloud and environmental scatter \citep{Utomo2018, Krumholz2019}. Other analyses argue that star formation efficiency evolves over the lifetime of a cloud and may therefore represent a time-averaged property of a rapidly evolving cloud population rather than a strict equilibrium constant \citep{Feldmann2010, Chevance2019}. Understanding how such low average efficiencies emerge from the interplay between turbulence, gravity, and feedback therefore remains an open theoretical problem.

In this work, we develop a minimal analytic framework for the dynamical evolution of turbulence in giant molecular clouds. Modeling each cloud as a one-zone turbulent system, we derive an ordinary differential equation for the velocity dispersion based on a balance between stellar feedback and turbulent dissipation, and demonstrate the existence of a stable low-efficiency equilibrium. We then extend the model by including gravity-driven turbulent driving to assess the robustness of this equilibrium. Finally, we compare the predictions with observed GMC properties and discuss the connection to cloud-scale and galaxy-averaged star formation relations.

The present model can be viewed as a dynamical realization of feedback-regulated star formation, providing a complementary perspective to turbulence-regulated models based on density statistics. Rather than introducing a new feedback mechanism, the framework emphasizes the time evolution and stability of cloud-scale turbulence, highlighting the leading-order interplay between turbulence, gravity, and stellar feedback.

\section{Momentum-Regulated Turbulence in GMCs}

In this section, we develop a momentum-based analytic framework to describe the turbulent state of giant molecular clouds. Stellar feedback injects momentum into the gas, driving turbulence that counteracts gravitational collapse. We show how this balance leads to a quasi-stable turbulent equilibrium associated with low star formation efficiencies, $\epsilon_{\rm ff}$.

Our model describes a single giant molecular cloud (GMC) as an effective one-zone system, in which global cloud-averaged quantities such as surface density $\Sigma$, size $L$, mass $M$, and velocity dispersion $\sigma$ characterize the dynamics. All equations and derived quantities, including the star formation efficiency per free-fall time $\epsilon_{\rm ff}$, refer to these cloud-averaged values rather than to local substructures. Galactic-scale relations emerge statistically when averaging over ensembles of clouds, each evolving toward its own self-regulated turbulent equilibrium.

\subsection{Feedback Driving and Turbulent Dissipation}

The effective momentum per unit stellar mass, $p_*/M_*$, sets the rate at which stellar feedback drives turbulence and counteracts gravitational collapse. It encapsulates contributions from supernovae, stellar winds, and radiation pressure, with typical values $p_*/M_* \sim 1000 - 3000$ km s$^{-1}$ depending on environment and stellar population \citep{Hopkins2012, Kim2021}. This range reflects both the diversity of star-forming environments and the relative contributions of different types of massive stars to the momentum budget. In this framework, the momentum injection rate can be expressed as
\begin{equation}
\dot{P}_{\rm fb} = \frac{p_*}{M_*} \dot{M}_*,
\end{equation}
where $\dot{M}_*$ is the instantaneous star formation rate. This relation expresses a direct negative feedback loop: higher star formation rates lead to stronger momentum injection and hence stronger turbulent support.

Turbulent motions in GMCs decay over roughly a crossing time, $t_{\rm cross} \sim L/\sigma$, where $L$ is the characteristic cloud size and $\sigma$ is the velocity dispersion \citep{Stone1998, MacLow2004}. The corresponding momentum dissipation rate can be estimated as
\begin{equation}
\dot{P}_{\rm diss} \sim \frac{M \sigma^2}{L},
\label{eq:dissipation}
\end{equation}
where $M$ is the cloud mass. The turbulent momentum scales as $M\sigma$ and decays over approximately one crossing time, $t_{\rm cross}\sim L/\sigma$. If dissipation is not balanced by feedback driving, turbulence decays and gravitational collapse accelerates. Conversely, strong turbulence suppresses collapse and reduces the star formation activity, establishing a self-regulating feedback loop.

\subsection{Momentum Balance and Star Formation Efficiency}

Assuming that the cloud reaches such a quasi-steady equilibrium, where feedback momentum injection approximately balances turbulent dissipation, $\dot{P}_{\rm fb} \approx \dot{P}_{\rm diss}$, we can relate the SFE per free-fall time, $\epsilon_{\rm ff}$, to the cloud properties:
\begin{equation}
\epsilon_{\rm ff} = \eta \,\frac{\dot{M}_* t_{\rm ff}}{M} \sim \eta\,\frac{\sigma^2 t_{\rm ff}}{L (p_*/M_*)}.
\end{equation}
Here, $t_{\rm ff}$ denotes the free-fall time of the cloud and $\eta \sim 5-10$ is an efficiency factor, which encapsulates the combined impact of clustered and overlapping stellar feedback, non-uniform density distributions, and geometric cloud structure \citep[see, e.g.,][]{Hopkins2012,Federrath2012,Agertz2013,Grudic2018}. It should be interpreted as an effective calibration parameter that absorbs unresolved cloud structure, clustered feedback, and imperfect coupling efficiencies. Values $\eta > 1$ reflect the enhanced momentum coupling expected in a highly inhomogeneous medium. In particular, clustered star formation and density fluctuations can enhance the effective coupling between feedback and gas, leading to a higher effective momentum input than predicted by uniform models \citep[e.g.,][]{Federrath2012, Grudic2018}. Constraining $\eta$ from simulations or resolved observations remains an important task for future work.

Throughout this paper we adopt the fiducial value $\eta=5$, corresponding to the lower end of the physically plausible range $\eta\sim5$--10. This choice represents moderately efficient momentum coupling between stellar feedback and the turbulent cloud gas and serves as a representative reference value for typical Milky Way GMC conditions. Larger values primarily rescale the normalization of the predicted star formation efficiency without changing the qualitative behavior of the model or the existence and stability of the self-regulated equilibrium.

Adopting the standard virial scaling, $\sigma^2 \sim GM/L$, and $t_{\rm ff} \sim \sqrt{L^3 / GM}$, the expression for the star formation efficiency per free-fall time can be written as
\begin{equation}
\epsilon_{\rm ff} 
\sim 
\eta
\frac{\sqrt{G \Sigma L}}{p_*/M_*}
\label{eq:epsilonff_estimate}
\end{equation}
where $\Sigma = M/L^2$ is the cloud surface density. This yields
\begin{equation}
\epsilon_{\rm ff}
\simeq
0.001
\eta
\left(\frac{\Sigma}{100\,M_\odot\,{\rm pc^{-2}}}\right)^{1/2}
\left(\frac{L}{20\,{\rm pc}}\right)^{1/2}
\left(\frac{p_*/M_*}{3000\,{\rm km\,s^{-1}}}\right)^{-1}.
\end{equation}

\subsection{Scaling Behavior}

Equation~\eqref{eq:epsilonff_estimate} implies a weak dependence of $\epsilon_{\rm ff}$ on surface density and cloud size. For typical GMC parameters ($L\sim10$–30 pc, $\Sigma \sim 50$–150 M$_\odot$ pc$^{-2}$), this yields $\epsilon_{\rm ff} \sim 0.5\% - 2\%$, consistent with observations \citep{Murray2011, Evans2014, Utomo2018}. This weak scaling arises because the turbulent velocity dispersion is itself dynamically regulated, partially compensating variations in cloud properties.

To assess the robustness of the analytic framework, we also explored the dependence of the predicted star formation efficiency on the principal free parameters of the model (see Appendix~\ref{appendix:model_parameters}). As expected from the analytic scaling relations, increasing $\eta$ raises the equilibrium efficiency, while increasing $p_*/M_*$ reduces it. These parameter variations shift the predicted efficiencies by factors of order unity, but do not alter the overall weak dependence of $\epsilon_{\rm ff}$ on gas surface density.

We also note that the quantity $\sqrt{G\Sigma L}$ appearing in the scaling relation corresponds to the virial velocity dispersion of a self-gravitating cloud. Using $M \sim \Sigma L^2$, the virial theorem gives
\begin{equation}
\sigma_{\rm vir}^2 \sim \frac{GM}{L} \sim G\Sigma L,
\end{equation}
so that $\sigma_{\rm vir} \sim \sqrt{G\Sigma L}$. Defining $\sigma_{\rm fb} = p_*/M_*$ as the feedback driven velocity dispersion, the equilibrium efficiency is
\begin{equation}
\epsilon_{\rm ff}
\sim
\eta
\frac{\sigma_{\rm vir}}{\sigma_{\rm fb}},
\end{equation}
demonstrating that the star formation efficiency is set by the ratio between the gravitational velocity scale of the cloud and the momentum injected per unit stellar mass by feedback. Table~\ref{tab:model_parameters} summarizes the principal model parameters and characteristic ranges adopted throughout this work. Unless otherwise stated, these values are chosen to represent typical conditions in Milky Way giant molecular clouds.

\begin{table}
\centering
\begin{tabular}{lll}
\hline
Quantity & Fiducial / Range & Role \\
\hline
$L$ & $10$--$30$ pc & Characteristic cloud size \\
$M$ & $10^4$--$10^6\,M_\odot$ & Cloud mass \\
$\Sigma$ & $50$--$150\,M_\odot\,{\rm pc}^{-2}$ & Cloud surface density \\
$\sigma$ & $1$--$5$ km s$^{-1}$ & Turbulent velocity dispersion \\
$t_{\rm ff}$ & $1.5$--$5$ Myr & Free-fall time of the cloud \\
$\eta$ & $5$--$10$ & Effective calibration factor \\
$p_*/M_*$ & $1000$--$3000$ km s$^{-1}$ & Injected momentum \\
$A$ & $(p_*/M_*)/(\eta t_{\rm ff})$ & Derived feedback coefficient \\
$B$ & $0$--$0.5$ & Gravity-driving coefficient \\
\hline
\end{tabular}
\vspace{0.5em}
\caption{Summary of the principal model parameters and characteristic ranges adopted throughout this work. The listed values are intended to represent typical Milky Way giant molecular cloud conditions and serve as fiducial inputs for the analytic estimates and numerical examples presented in the following sections.}
\label{tab:model_parameters}
\end{table}

\section{Turbulent Evolution and Equilibrium Stability}

In this section, we formulate a simple ordinary differential equation for the turbulent velocity dispersion, allowing us to analyze the relaxation toward equilibrium and the stability of the feedback-regulated state in a transparent analytic way.

\subsection{ODE for Turbulent Evolution}

At a qualitative level, the evolution of the turbulent velocity dispersion can be written as
\begin{equation}
\frac{d\sigma}{dt} \sim \text{Feedback} - \text{Dissipation},
\end{equation}
where ``Feedback'' represents the momentum per unit mass injected by young stars, and ``Dissipation'' captures the decay of turbulence over roughly a crossing time.

Let $\sigma$ denote the turbulent velocity dispersion of a cloud. The time evolution of $\sigma$ can then be approximated as
\begin{equation}
\frac{d\sigma}{dt} = \frac{\dot{P}_{\rm fb}}{M} - \frac{\sigma}{t_{\rm diss}},
\end{equation}
where $\dot{P}_{\rm fb} = (p_*/M_*) \dot{M}_*$ is the momentum injection rate per unit cloud mass, and $t_{\rm diss} \sim L/\sigma$ is the turbulent dissipation timescale.
From this it follows that
\begin{equation}
\frac{d\sigma}{dt} = \frac{(p_*/M_*) \dot{M}_*}{M} - \frac{\sigma^2}{L},
\end{equation}
which serves as the basis for our analytic equilibrium and stability analysis. Substituting the expressions for $\epsilon_{\rm ff}$, the ODE becomes
\begin{equation}
\frac{d\sigma}{dt} = \frac{(p_*/M_*) \epsilon_{\rm ff}}{\eta t_{\rm ff}} - \frac{\sigma^2}{L}.
\label{eq:sigma_ODE}
\end{equation}
The resulting evolution equation captures the competition between feedback driving and turbulent dissipation in the simplest dynamical form consistent with the one-zone approximation. However, we note that in the fiducial model considered here, $\epsilon_{\rm ff}$ is treated as an externally specified effective parameter, allowing us to isolate the dynamical response of the turbulent velocity dispersion. A fully self-consistent model, in which $\epsilon_{\rm ff}$ is coupled to the turbulent state, is introduced in Section~4.

Setting $d\sigma/dt = 0$, the equilibrium velocity dispersion can be easily computed and is
\begin{equation}
\sigma_{\rm eq} = \sqrt{\frac{(p_*/M_*) \epsilon_{\rm ff} L}{\eta t_{\rm ff}}}.
\end{equation}
This is the velocity dispersion at which momentum injection balances turbulent dissipation, corresponding to a quasi-steady state of the cloud, consistent with our result presented in Eq.~\eqref{eq:epsilonff_estimate}.

\subsection{Stability Analysis of the Quasi-Stable Equilibrium}

To test the equilibrium stability, we now consider a small perturbation $\delta \sigma(t)$ around the equilibrium point $\sigma_{\rm eq}$:
\begin{equation}
\sigma(t) = \sigma_{\rm eq} + \delta \sigma(t), \quad |\delta \sigma| \ll \sigma_{\rm eq}.
\label{eq:sigma_perturb}
\end{equation}
Linearizing the ODE for $\delta \sigma$ by plugging Eq.~\eqref{eq:sigma_perturb} into Eq.~\eqref{eq:sigma_ODE} and neglecting higher-order terms, we find
\begin{equation}
\frac{d (\delta \sigma)}{dt} = - \frac{2 \sigma_{\rm eq}}{L} \, \delta \sigma,
\end{equation}
where $\delta \sigma = \sigma - \sigma_{\rm eq}$ represents a small perturbation from the equilibrium. We can define the relaxation timescale $\tau_{\rm relax}$ as
\begin{equation}
\tau_{\rm relax} \equiv \frac{L}{2 \sigma_{\rm eq}}.
\end{equation}
This timescale characterizes how quickly the system returns to equilibrium after a perturbation. The solution to the linearized equation is then
\begin{equation}
\delta \sigma(t) = \delta \sigma(0) \, \exp\left(-\frac{t}{\tau_{\rm relax}}\right),
\end{equation}
demonstrating that deviations from equilibrium decay exponentially. Physically, $\tau_{\rm relax}$ is of order the cloud crossing time divided by a factor of 2, meaning that any temporary excess or deficit in turbulence is rapidly damped over roughly a few crossing times. The equilibrium is therefore linearly stable, with perturbations decaying on the relaxation timescale. Even moderate perturbations in density, feedback strength, or star formation rate do not destabilize the cloud; they are damped on a timescale $\tau_{\rm relax}$. Consequently, the ODE framework not only reproduces the quasi-steady turbulent velocity but also provides a quantitative measure of the speed of relaxation toward the equilibrium.

To quantify the sensitivity of the star formation efficiency to variations in cloud properties, we consider the total differential of $\epsilon_{\rm ff}$. Using the scaling from above,
\begin{equation}
\epsilon_{\rm ff} \propto 
\eta \frac{\sigma^2}{p_*/M_*}
\Sigma^{-1/2}
L^{-1/2},
\end{equation}
we obtain
\begin{equation}
\frac{\delta \epsilon_{\rm ff}}{\epsilon_{\rm ff}}
=
\frac{\delta \eta}{\eta}
+ 2 \frac{\delta \sigma}{\sigma}
- \frac{\delta (p_*/M_*)}{p_*/M_*}
- \frac{1}{2} \frac{\delta \Sigma}{\Sigma}
- \frac{1}{2} \frac{\delta L}{L}.
\end{equation}
This expression shows that $\epsilon_{\rm ff}$ depends only weakly on global cloud properties such as surface density and size, while variations in $\sigma$ are dynamically suppressed by the self-regulating feedback loop. As a result, the star formation efficiency tends to cluster around a narrow range despite variations in cloud conditions. A more quantitative assessment of these parameter sensitivities is presented in Appendix~\ref{appendix:SFE_sensitivity}.

\subsection{Relevant Timescales in our ODE}

The free-fall time of a cloud is
\begin{equation}
t_{\rm ff} = \sqrt{\frac{3 \pi}{32 G \rho}}
\approx
4.4~{\rm Myr}
\left(
\frac{n_{\rm H}}{100~{\rm cm^{-3}}}
\right)^{-1/2},
\end{equation}
where $\rho$ is the mean cloud density and $n_{\rm H}$ the corresponding hydrogen number density. For typical GMC conditions with
$n_{\rm H} \sim 50$--500~cm$^{-3}$, this implies $t_{\rm ff} \sim 1.5\text{--}5~{\rm Myr}$. Using representative cloud parameters
$L \sim 10$--50~pc and
$\sigma_{\rm eq} \sim 1$--5~km~s$^{-1}$,
the relaxation timescale becomes
\begin{equation}
\tau_{\rm relax}
=
\frac{L}{2\sigma_{\rm eq}}
\sim
1\text{--}5~{\rm Myr},
\end{equation}
i.e., of the same order as the cloud free-fall time. The comparable magnitudes of
$\tau_{\rm relax}$ and $t_{\rm ff}$ imply that turbulence adjusts on approximately dynamical timescales, allowing GMCs to evolve toward quasi-stable turbulent states during their evolution. Because
$\tau_{\rm relax} \propto L/\sigma_{\rm eq}$,
larger clouds or systems with lower equilibrium velocity dispersions relax more slowly, producing moderate cloud-to-cloud variations in the resulting star formation efficiency.

\subsection{Model Assumptions}
\label{subsec:model_assumptions}

Our analytic framework is intentionally minimal and describes the turbulent evolution of GMCs using a low-dimensional dynamical model based on a small set of simplifying assumptions. We assume that GMCs evolve toward a quasi-steady turbulent equilibrium characterized by
\begin{equation}
\frac{d\sigma}{dt}
\approx
\dot{\sigma}_{\rm drive}
-
\dot{\sigma}_{\rm diss}
\approx 0,
\end{equation}
implying $\dot{\sigma}_{\rm drive} \simeq \dot{\sigma}_{\rm diss}$. This is motivated by the fact that the turbulent crossing time, $t_{\rm cross} \sim L/\sigma$, is typically comparable to cloud evolutionary timescales, allowing turbulence to relax efficiently toward equilibrium, although early assembly or dispersal phases may deviate from this behavior. Furthermore, stellar feedback is assumed to inject momentum that couples efficiently to the gas and drives turbulence,
\begin{equation}
\dot{\sigma}_{\rm fb}
\sim
\frac{p_*}{M_*}\,
\frac{\dot{M}_*}{M_{\rm cloud}},
\end{equation}
where $\dot{M}_* = \epsilon_{\rm ff} M_{\rm cloud}/t_{\rm ff}$. Different feedback channels are subsumed into an effective momentum budget, with uncertainties absorbed into normalization parameters. Beyond this, turbulence is assumed to be driven and dissipated on a scale comparable to the cloud size $L$, with dissipation approximated as $\dot{\sigma}_{\rm diss} \sim \sigma^2/L$, corresponding to decay over approximately one crossing time. While real turbulence spans a range of scales, the largest scales dominate the energy budget, justifying the single-scale approximation. Finally, magnetic fields are neglected in the fiducial model. To leading order, their effect can be represented as an additional contribution to the effective velocity dispersion,
$\sigma_{\rm eff}^2 = \sigma^2 + v_A^2$, which primarily shifts the equilibrium normalization without qualitatively altering the self-regulated turbulent state.

\section{Extended Dynamical Model: Gravitational Driving and Turbulent Self-Regulation}

In the fiducial model discussed above, turbulent motions are regulated by a balance between stellar feedback and turbulent dissipation. However, gravitational collapse itself can also contribute to the turbulent energy budget through converging flows and hierarchical contraction within molecular clouds. Numerical simulations and analytic arguments suggest that collapse-driven motions may replenish turbulence on approximately a free-fall timescale \citep{Robertson2012, Murray2015, Krumholz2018}. To account for this effect, we introduce an additional gravity-driven turbulent driving term into the dynamical evolution equation. In contrast to the equilibrium scaling derived in the previous section, we now treat $\epsilon_{\rm ff}$ as a dynamical quantity to enable a self-consistent time evolution.

\subsection{Extended Turbulent Evolution Equation}

The turbulent velocity dispersion evolves through a competition between driving and dissipation,
\begin{equation}
\frac{d\sigma}{dt}
=
\dot{\sigma}_{\rm fb}
+
\dot{\sigma}_{\rm grav}
-
\dot{\sigma}_{\rm diss},
\end{equation}
where the three terms represent feedback driving, gravity-driven turbulence, and turbulent dissipation, respectively. Stellar feedback injects momentum through supernovae, stellar winds, and radiation pressure, acting as an external source of turbulent support. At the same time, gravitational collapse can generate turbulent motions through converging flows and hierarchical contraction, effectively converting gravitational potential energy into turbulent kinetic energy. Opposing both driving mechanisms, supersonic turbulence decays through shocks and turbulent cascades on approximately a cloud crossing time, leading to dissipation proportional to $\sigma^2/L$.

Taking these effects into account, the evolution equation for the turbulent velocity dispersion extends Eq.~\eqref{eq:sigma_ODE} and becomes
\begin{equation}
\frac{d\sigma}{dt}
=
A\,\epsilon_{\rm ff}
+
B\frac{\sigma}{t_{\rm ff}}
-
\frac{\sigma^2}{L},
\label{eq:new_sigma}
\end{equation}
where the first term represents feedback-driven momentum injection, the second term parametrizes gravity-driven turbulent amplification, and the final term describes turbulent dissipation over approximately one crossing time. Here,
\begin{equation}
A
=
\frac{p_*/M_*}{\eta t_{\rm ff}},
\label{eq:Aconstant}
\end{equation}
while the dimensionless coefficient $B$ characterizes the efficiency with which gravitational contraction feeds turbulent motions. Furthermore, $\epsilon_{\rm ff}$ should be interpreted as the effective star formation efficiency emerging from the underlying momentum balance discussed in the fiducial model. The present formulation therefore focuses on the dynamical response of the turbulent velocity dispersion, while $\epsilon_{\rm ff}$ is treated as an effective quantity that can be coupled to the turbulent state.

We note that the structure of Eq.~\eqref{eq:new_sigma}, in which the evolution of the turbulent velocity dispersion is governed by the competition between feedback, gravitational driving, and dissipation, is conceptually similar to energy balance models developed for galactic discs \citep[e.g.,][]{Krumholz2018, Ginzburg2022, Forbes2023}. In these works, turbulence is driven both by stellar feedback and by gravitational energy released through large-scale processes such as radial mass transport and gravitational instability.

In contrast, the present model is formulated at the scale of individual molecular clouds and does not attempt to model the detailed physical origin of the gravitational driving term explicitly. Instead, the term $B\,\sigma/t_{\rm ff}$ should be interpreted as a phenomenological parametrization of turbulence generated by collapse-driven motions, such as converging flows and hierarchical contraction within self-gravitating gas. Thus, $B$ measures the efficiency with which gravitational potential energy is converted into turbulent motions on cloud scales. The linear dependence on $\sigma$ is chosen as the simplest closure consistent with a one-zone description; other scalings, for example $\propto \sigma^2/t_{\rm ff}$, would be equally plausible in more detailed models and would mainly rescale the equilibrium relation without changing the qualitative self-regulation picture. In the limit $B\to 0$, the model reduces to the fiducial feedback-regulated case discussed above. Larger values of $B$ are expected in clouds where collapse-driven flows contribute substantially to the turbulent energy budget, for example in more strongly self-gravitating or dynamically evolving systems. Accordingly, $B$ should be regarded as an effective environment-dependent parameter rather than a universal constant.

Setting our ODE to zero, i.e. $d\sigma/dt = 0$, the equilibrium velocity dispersion satisfies
\begin{equation}
A\,\epsilon_{\rm ff}
+
B\frac{\sigma_{\rm eq}}{t_{\rm ff}}
=
\frac{\sigma_{\rm eq}^2}{L}.
\label{eq:extended_equilibrium}
\end{equation}
Equation~\eqref{eq:extended_equilibrium} shows explicitly that the equilibrium turbulent state is determined by the balance between feedback-driven momentum injection, gravity-driven turbulent amplification, and turbulent dissipation.

\subsection{Feedback- and Gravity-Influenced Regimes}

Although the model does not predict completely separate dynamical phases, it is nevertheless useful to distinguish between limiting physical regimes in which either stellar feedback or gravitational contraction contributes more strongly to the turbulent driving budget.

In feedback-dominated clouds, momentum injection from young stars provides the dominant source of turbulent support. Turbulence is replenished primarily by stellar feedback, while turbulent dissipation continuously removes kinetic energy. In this limit, the equilibrium condition approximately reduces to
\begin{equation}
A\,\epsilon_{\rm ff}
\simeq
\frac{\sigma_{\rm eq}^2}{L},
\end{equation}
recovering the fiducial momentum-regulated equilibrium discussed in Eq.~\eqref{eq:epsilonff_estimate}.

In contrast, gravity-influenced clouds may experience substantial turbulent amplification directly from gravitational contraction and converging flows. In this case, the gravity-driven term contributes significantly to maintaining the turbulent velocity dispersion, and the equilibrium condition becomes approximately
\begin{equation}
B\frac{\sigma_{\rm eq}}{t_{\rm ff}}
\simeq
\frac{\sigma_{\rm eq}^2}{L}.
\end{equation}
Solving for the equilibrium velocity dispersion yields
\begin{equation}
\sigma_{\rm eq}
\simeq
B\frac{L}{t_{\rm ff}}.
\end{equation}
Using the free-fall scaling
\begin{equation}
t_{\rm ff}
\propto
\rho^{-1/2}
\sim
\left(\frac{\Sigma}{L}\right)^{-1/2},
\end{equation}
the equilibrium velocity dispersion becomes
\begin{equation}
\sigma_{\rm eq}
\propto
B\sqrt{\Sigma L},
\end{equation}
which is comparable to the virial velocity scaling expected for self-gravitating clouds.

The distinction between these regimes should therefore not be interpreted as a sharp physical transition, but rather as a shift in the relative importance of different turbulent driving mechanisms. Real GMCs likely occupy intermediate states in which both stellar feedback and gravitational contraction contribute simultaneously to the turbulent energy budget.

\subsection{Closure Relation for the Star Formation Efficiency}

The evolution equation for the turbulent velocity dispersion (Eq.~\ref{eq:sigma_ODE}) depends explicitly on the SFE per free-fall time, $\epsilon_{\rm ff}$. The scaling $\epsilon_{\rm ff} \propto \sigma^2$ derived in the fiducial momentum-regulated model follows from the equilibrium condition and therefore does not provide an independent closure for the dynamical evolution.

To construct a self-consistent dynamical model, an explicit functional dependence $\epsilon_{\rm ff}(\sigma)$ must be specified. Physically, strong turbulence is expected to suppress gravitational collapse, implying that $\epsilon_{\rm ff}$ cannot increase indefinitely with $\sigma$. Instead, a negative coupling between turbulence and star formation efficiency must emerge at sufficiently large velocity dispersions. In the following, $\epsilon_{\rm ff}$ is thus treated as a dynamical quantity coupled to the turbulent state through Eq.~\eqref{eq:epsilon_closure}, replacing the fixed-efficiency assumption used in the fiducial model. To capture this behavior, we adopt a simple interpolating form,
\begin{equation}
\epsilon_{\rm ff}(\sigma)
=
\epsilon_0
\left(
\frac{\sigma}{\sigma_0}
\right)^2
\left[
1 +
\left(
\frac{\sigma}{\sigma_0}
\right)^3
\right]^{-1},
\label{eq:epsilon_closure}
\end{equation}
where $\sigma_0$ is a characteristic velocity scale, which we take to be of order the virial velocity dispersion, and $\epsilon_0$ is a normalization constant. This expression reproduces the momentum-regulated scaling $\epsilon_{\rm ff} \propto \sigma^2$ in the limit $\sigma \ll \sigma_0$, while introducing a suppression $\epsilon_{\rm ff} \propto \sigma^{-1}$ at large $\sigma \gg \sigma_0$. The resulting turnover ensures that excessive turbulence reduces the SFE, providing the negative feedback required for a stable equilibrium solution.

We emphasize that Eq.~\eqref{eq:epsilon_closure} should be interpreted as a phenomenological closure relation rather than a fundamental physical law. The qualitative dynamical behavior of the system, in particular the existence of a stable equilibrium, does not depend sensitively on the specific functional form adopted, provided that the closure introduces a negative coupling between turbulence and star formation efficiency at large $\sigma$. We tested several alternative forms, including a saturating relation and a logistic turnover, and found that the existence and stability of the fixed point remain unchanged (see Appendix~\ref{appendix:sensitivity}). Equation~(\ref{eq:epsilon_closure}) is therefore adopted as a representative minimal ansatz.

The adopted functional form is motivated by the expectation that weak turbulence allows collapse to proceed, whereas strong turbulence suppresses gravitational fragmentation and reduces the star formation efficiency. Similar non-monotonic dependencies are found in turbulence-regulated star formation models based on density statistics \citep[see, e.g.,][]{Federrath2012,Burkhart2018}.

\subsection{Linear Stability of the Extended Equilibrium}

The linear stability of the extended equilibrium solution can be analyzed in analogy with the fiducial model discussed above. Let
\begin{equation}
\sigma(t)
=
\sigma_{\rm eq}
+
\delta\sigma(t),
\end{equation}
where $|\delta\sigma| \ll \sigma_{\rm eq}$ denotes a small perturbation around the equilibrium state. Linearizing Eq.~\eqref{eq:new_sigma} around $\sigma_{\rm eq}$ yields
\begin{equation}
\frac{d}{dt}(\delta\sigma)
=
f'(\sigma_{\rm eq})\,\delta\sigma,
\end{equation}
with
\begin{equation}
f'(\sigma)
=
\frac{B}{t_{\rm ff}}
-
\frac{2\sigma}{L}.
\end{equation}
For equilibrium states satisfying
\begin{equation}
\frac{2\sigma_{\rm eq}}{L}
>
\frac{B}{t_{\rm ff}},
\end{equation}
the dissipation term dominates the linear response and
\begin{equation}
f'(\sigma_{\rm eq}) < 0,
\end{equation}
so that perturbations decay exponentially with time. Physically, excess turbulence is damped by dissipation, while insufficient turbulence enhances the relative importance of the driving terms, restoring the system toward equilibrium.

An equivalent stability argument follows from the Lyapunov function
\begin{equation}
V(\sigma)
=
\frac12
(\sigma-\sigma_{\rm eq})^2,
\end{equation}
whose time derivative becomes
\begin{equation}
\frac{dV}{dt}
=
(\sigma-\sigma_{\rm eq})f(\sigma).
\end{equation}
For stable equilibria, deviations from $\sigma_{\rm eq}$ are continuously damped by the competition between turbulent driving and dissipation, implying
\begin{equation}
\frac{dV}{dt}<0
\qquad
{\rm for}
\qquad
\sigma \neq \sigma_{\rm eq}.
\end{equation}
We note that for a one-dimensional autonomous system such as Eq.~\eqref{eq:new_sigma}, the existence of a Lyapunov function of this form is not unexpected. The purpose of this construction is therefore not to establish mathematical novelty, but to make the stability properties of the momentum-regulated equilibrium explicit and to provide a physically transparent interpretation of the relaxation toward equilibrium. The turbulent state therefore behaves as a dynamically stable equilibrium within the simplified one-zone framework considered here. Additional gravitational driving modifies the equilibrium structure quantitatively, but the qualitative self-regulating behavior of the system is preserved.

\begin{figure*}
\centering
\includegraphics[width=0.9\textwidth]{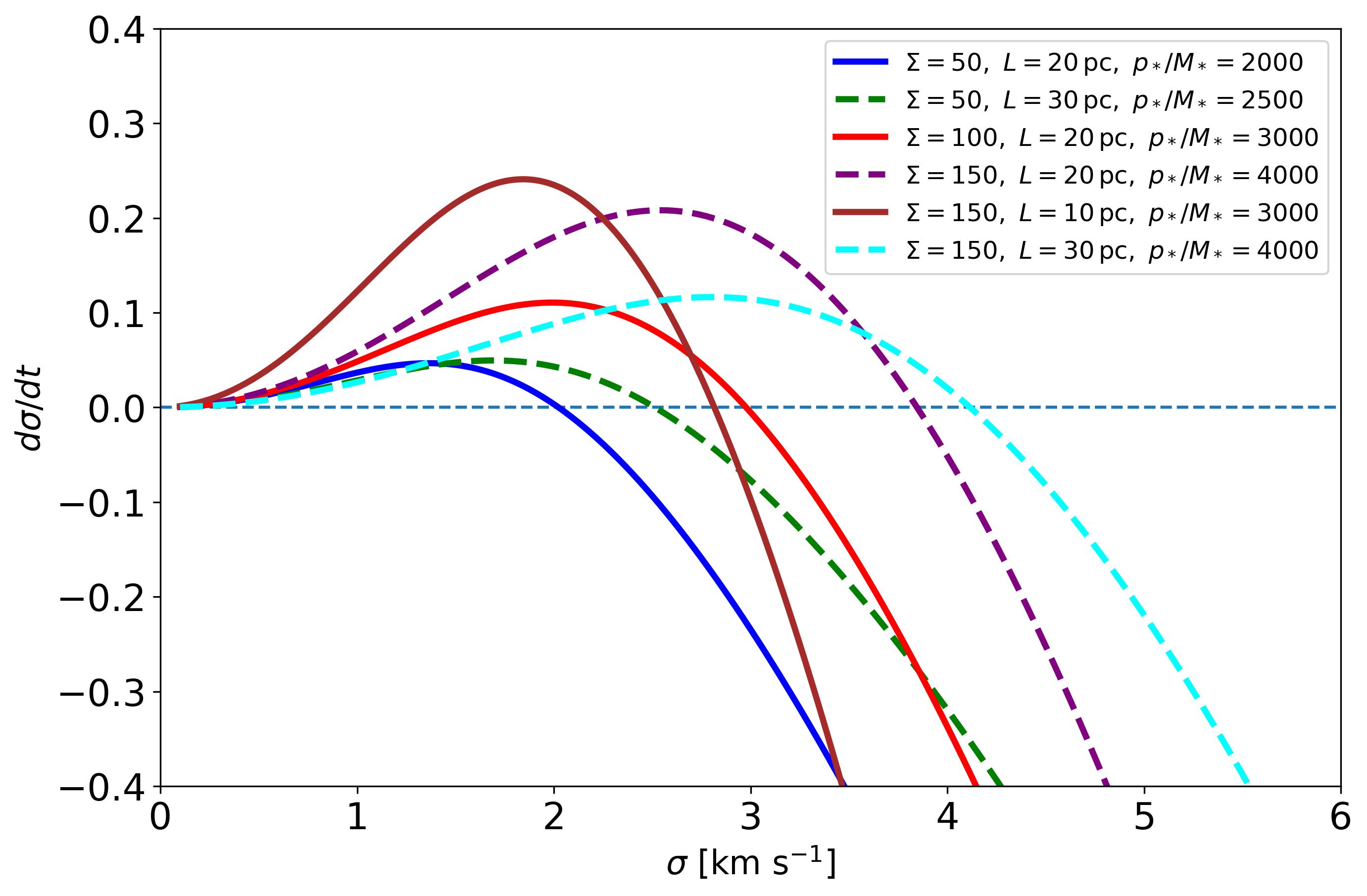}
\caption{Phase portrait of the turbulent velocity evolution, showing $d\sigma/dt$ as a function of $\sigma$ for six representative cloud models with varying surface densities $\Sigma$, sizes $L$, and feedback strengths $p_*/M_*$ for $\eta = 5$. The zero-crossing of each curve defines a stable equilibrium velocity dispersion. The sign of $d\sigma/dt$ demonstrates that the system evolves toward this equilibrium from both lower and higher turbulence levels, illustrating the presence of a dynamically stable equilibrium. The equilibrium value shifts moderately with cloud parameters, but the qualitative behavior remains robust.}
\label{fig:sigma_phase_portrait}
\end{figure*}

To further illustrate the dynamical behavior of the system, we show in Figure~\ref{fig:sigma_phase_portrait} the phase portrait of the turbulent velocity evolution, i.e., $d\sigma/dt$ as a function of $\sigma$, for a set of representative cloud models with a value of $\eta = 5$. Varying $\eta$ within the plausible range $\sim 5$--10 primarily rescales the equilibrium position but does not alter the qualitative behavior of the phase portrait or the existence of a stable equilibrium. Each curve corresponds to a different combination of cloud surface density, size, and feedback strength. The zero-crossing of each curve defines an equilibrium velocity dispersion $\sigma_{\rm eq}$. For all models considered, the system exhibits $d\sigma/dt > 0$ at low $\sigma$ and $d\sigma/dt < 0$ at high $\sigma$, demonstrating that the equilibrium is dynamically stable.

The phase portrait is constructed using the closure relation for the star formation efficiency introduced in Eq.~\eqref{eq:epsilon_closure}, which provides a self-consistent coupling between turbulence and star formation. In particular, the turnover in $\epsilon_{\rm ff}(\sigma)$ at large $\sigma$ ensures that the feedback driving term does not grow indefinitely, enabling the existence of a stable fixed point. The phase portrait therefore provides a direct visualization of the self-regulating equilibrium described above. Importantly, the existence of this equilibrium is robust across a wide range of cloud parameters, with only moderate shifts in the equilibrium value. This suggests that the self-regulation mechanism operates generically in typical GMC conditions.

\subsection{Robustness of the Self-Regulated Equilibrium}

A natural question is whether additional physical processes omitted from the fiducial feedback--dissipation model could significantly alter the predicted equilibrium SFE, $\epsilon_{\rm ff}$. In the following, we briefly discuss several important candidates.

Radiation pressure from massive stars injects additional momentum into the surrounding gas and may contribute to turbulent support. Numerical simulations suggest that in typical Milky Way GMCs, its contribution is smaller than that of supernovae and stellar winds, although it may become significant in dense star-forming regions \citep{Murray2011, KimOstriker2015}. Within our framework, this effect primarily rescales the effective feedback coefficient $A$ rather than introducing a qualitatively new dynamical timescale.

Magnetic fields likewise provide additional support against gravitational compression. To leading order, magnetic pressure can be represented through an effective velocity dispersion, $\sigma_{\rm eff}^2=\sigma^2+v_A^2$, where $v_A$ is the Alfv\'en speed. Observationally inferred Alfv\'en speeds in GMCs are typically comparable to, but not larger than, turbulent velocities \citep{Crutcher2012,HennebelleInutsuka2019}, implying that magnetic fields mainly shift the equilibrium normalization without qualitatively changing the quasi-stable equilibrium behavior.

Additional sources of pressure support, including cosmic rays and thermal pressure from warm or hot gas phases, evolve on characteristic timescales longer than the turbulent crossing time. Consequently, these components primarily modify the global cloud structure while only weakly affecting the rapid turbulent relaxation described by Eq.~\eqref{eq:new_sigma}. Collectively, these additional physical processes primarily modify the coefficients entering the dynamical evolution equation rather than removing the existence of a stable equilibrium solution.

\subsection{Physical Interpretation}

The momentum-regulated framework developed suggests that the low SFEs observed in GMCs arise from a dynamical interplay between stellar feedback, turbulent dissipation, and gravitational processes. Stellar feedback continuously injects momentum into the gas, replenishing turbulence and counteracting gravitational collapse, while turbulent motions dissipate on a crossing time. This interplay establishes a self-regulating feedback loop. If star formation temporarily increases, enhanced feedback strengthens the turbulent velocity dispersion, suppressing further collapse. Conversely, if turbulence becomes too strong, star formation and therefore feedback decline, allowing turbulence to decay and gravity to regain influence. As a result, the system evolves toward a quasi-stable turbulent state in which momentum injection and dissipation remain approximately balanced.

The inclusion of gravity-driven turbulent amplification modifies this picture quantitatively, but does not alter the underlying regulatory mechanism. Collapse-driven motions contribute additional turbulent energy, effectively shifting the equilibrium velocity dispersion. However, the system retains its self-regulating character, with deviations from equilibrium damped by the competition between driving and dissipation.

Within this unified framework, gravitational driving primarily adjusts the normalization of the turbulent state rather than introducing qualitatively new dynamics. Consequently, the emergence of low star formation efficiencies can be understood as a robust outcome of turbulent self-regulation, largely independent of whether the dominant driving source is stellar feedback or gravitational contraction. Moderate variations in cloud properties therefore lead only to correspondingly modest changes in $\epsilon_{\rm ff}$, consistent with the limited scatter observed in typical Milky Way GMCs.

\begin{figure*}
\centering
\includegraphics[width=0.9\textwidth]{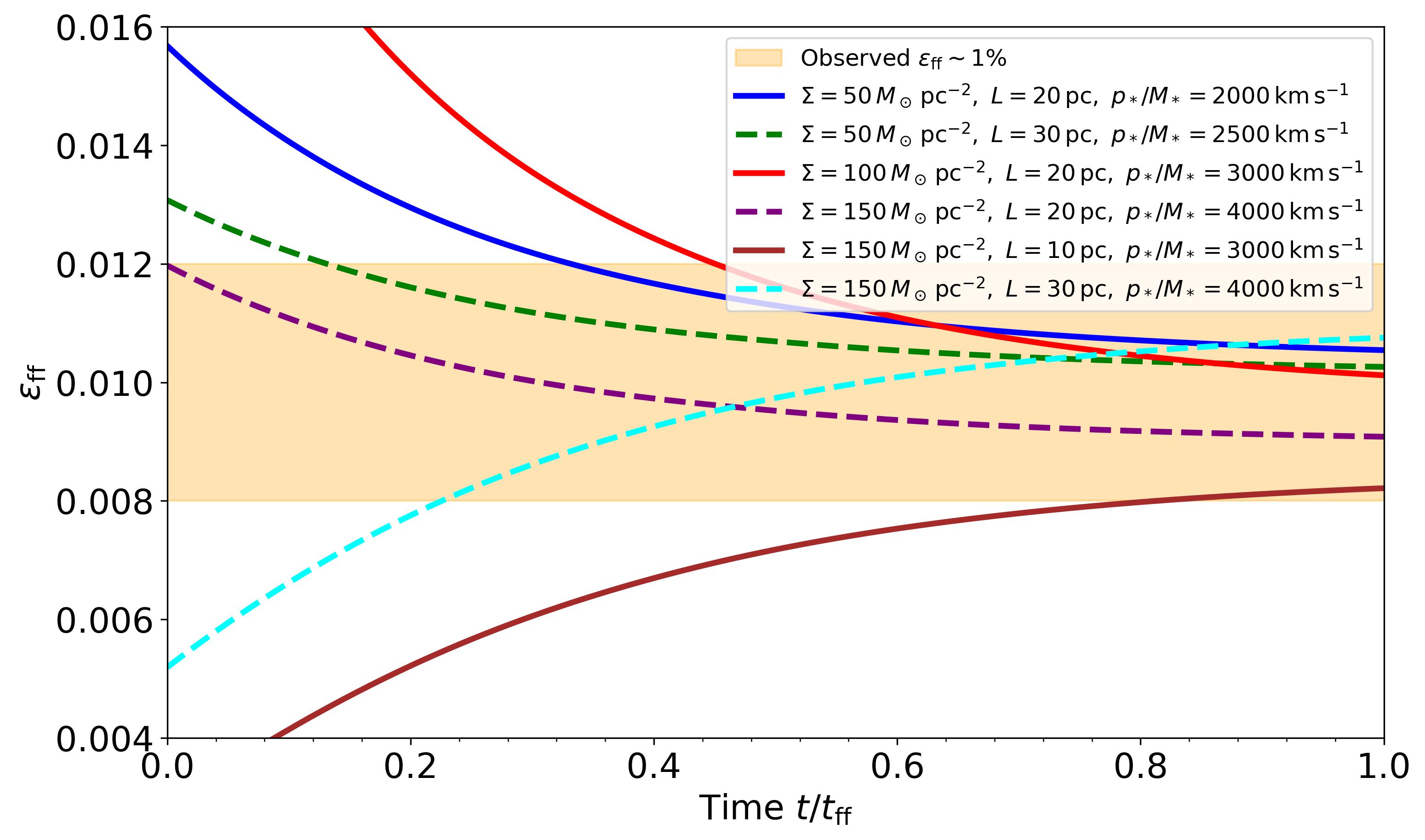}
\caption{Emergent star formation efficiencies, $\epsilon_{\rm ff}$, as a function of time normalized by the cloud free-fall time, $t/t_{\rm ff}$, for six representative molecular cloud models. Each curve corresponds to a cloud with specific surface density $\Sigma$, size $L$, $\eta = 5$, and momentum injection per unit stellar mass $p_*/M_*$, as indicated in the legend. The figure illustrates the dynamic evolution of $\epsilon_{\rm ff}$ from an initially turbulent state toward a quasi-stable equilibrium determined by the balance between feedback-driven momentum injection and turbulent dissipation. The shaded orange region marks the observational range between $\epsilon_{\rm ff} \sim 0.01 \pm 0.005$. All six cloud models converge to values within or near this observational band, demonstrating that our momentum-regulated framework produces the limited variation of star formation efficiencies seen in real GMCs, while allowing for modest variation due to differences in cloud surface density, size, and feedback strength. This highlights the robustness of the model and captures key physical processes influencing $\epsilon_{\rm ff}$ across diverse cloud environments.}
\label{fig:epsilon_ff_dynamic}
\end{figure*}

\begin{figure*}
\centering
\includegraphics[width=0.9\textwidth]{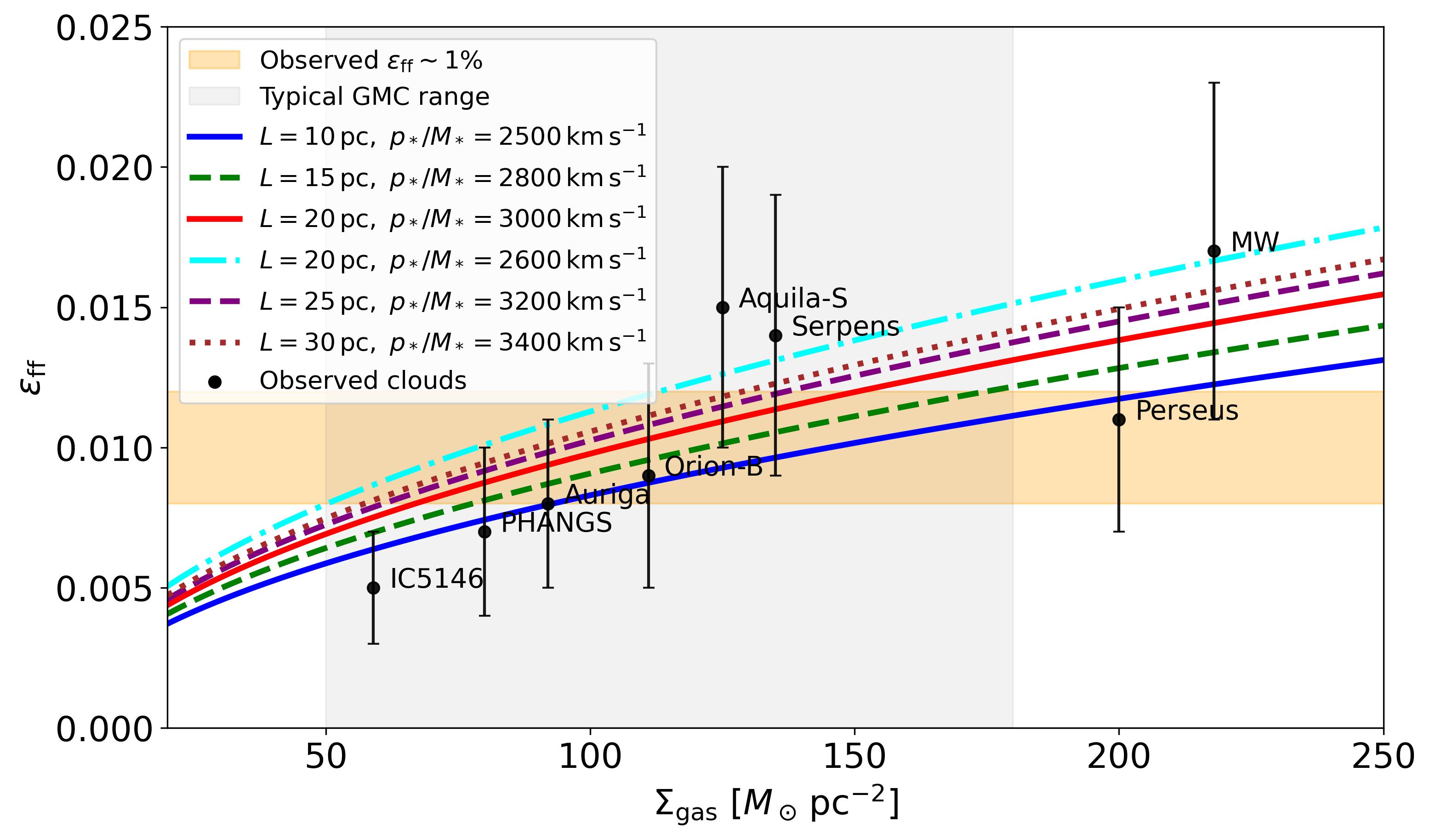}
\caption{Predicted star formation efficiency per free-fall time, $\epsilon_{\rm ff}$, as a function of gas surface density, $\Sigma_{\rm gas}$, for a set of representative cloud models with varying sizes $L$ and momentum injection efficiencies $p_*/M_*$. The shaded orange region indicates the typical observational range, while the grey band marks the characteristic surface density range of Milky Way giant molecular clouds ($\Sigma \sim 50$--180 $M_\odot\,\mathrm{pc}^{-2}$). The normalization depends on the adopted value of $\eta$ (here we use $\eta = 5$), but the weak dependence on $\Sigma$ remains unchanged. The model predicts a weak increase of $\epsilon_{\rm ff}$ with surface density, but over the limited range occupied by typical GMCs the efficiency remains approximately constant at the $\sim 1\%$ level. Outside this regime, systematic deviations emerge, with lower efficiencies at low surface densities and higher efficiencies at large $\Sigma$, reflecting the limits of the momentum-regulated equilibrium in more diffuse or extreme environments. The black points with error bars show observed molecular clouds from Table~\ref{tab:gmc_comparison}. The data lie within the region spanned by the model curves and cluster around the characteristic efficiency, with a scatter comparable to the observational uncertainties.}
\label{fig:epsilon_ff_sigma}
\end{figure*}

\section{The Kennicutt--Schmidt Relation Revisited}

The Kennicutt--Schmidt (KS) relation empirically links the surface density of star formation, $\Sigma_{\rm SFR}$, to the gas surface density, $\Sigma_{\rm gas}$, typically as a power law:
\begin{equation}
\Sigma_{\rm SFR} \propto \Sigma_{\rm gas}^{N}.
\end{equation}
Historically, \citet{Schmidt1959} found $N \sim 2$ from Milky Way star formation regions, whereas \citet{Kennicutt1998} reported a somewhat shallower slope of $N \sim 1.4-1.5$ when averaging over entire galaxies. In this section, we demonstrate how this relation emerges from the momentum-regulated dynamical model developed above. Rather than assuming a star formation law \emph{a priori}, we derive the scaling between $\Sigma_{\rm SFR}$ and $\Sigma_{\rm gas}$ from the equilibrium between gravitational driving, turbulent dissipation, and stellar feedback. In particular, we show how different effective behaviors of the star formation efficiency, arising from local equilibrium versus population averaging, lead to the classical Schmidt ($N\simeq2$) and Kennicutt ($N\simeq1.5$) slopes, thereby connecting cloud-scale physics to galaxy-scale star formation laws \citep{Kennicutt2012}.

\subsection{Relation to Momentum-Regulated Star Formation}

In our model, the star formation surface density is given by
\begin{equation}
\Sigma_{\rm SFR}
=
\epsilon_{\rm ff}
\frac{\Sigma_{\rm gas}}{t_{\rm ff}},
\label{eq:sfr_general}
\end{equation}
where $\epsilon_{\rm ff}$ is the star formation efficiency per free-fall time and $t_{\rm ff}$ is the gravitational free-fall time. For a cloud of characteristic size $L$, the mean density scales as
\begin{equation}
\rho \sim \frac{\Sigma_{\rm gas}}{L}.
\end{equation}
The free-fall time therefore becomes
\begin{equation}
t_{\rm ff}
\propto
\rho^{-1/2}
\propto
\left(\frac{\Sigma_{\rm gas}}{L}\right)^{-1/2}.
\end{equation}
Substituting into Eq.~\eqref{eq:sfr_general} gives
\begin{equation}
\Sigma_{\rm SFR}
\propto
\epsilon_{\rm ff}\,
\Sigma_{\rm gas}^{3/2}
L^{-1/2}.
\label{eq:sfr_master}
\end{equation}
Equation~(\ref{eq:sfr_master}) provides the general scaling from which different KS slopes emerge depending on the behavior of $\epsilon_{\rm ff}$. We will discuss this in the following sections.

\subsection{Cloud-scale Scaling: Schmidt Law ($N=2$)}

On the scale of individual clouds, the turbulence adjusts dynamically to balance feedback and dissipation, setting a local equilibrium. The resulting star formation efficiency is therefore sensitive to the internal velocity dispersion, which reflects both the cloud's mass and size. In the momentum-regulated framework developed above, the equilibrium star formation efficiency follows from the balance between feedback momentum injection and turbulent dissipation, yielding
\begin{equation}
\epsilon_{\rm ff}
\propto
\frac{\sigma^2 t_{\rm ff}}{L (p_*/M_*)}.
\end{equation}
Using the virial scaling $\sigma^2 \sim G \Sigma_{\rm gas} L$ and $t_{\rm ff} \propto (\Sigma_{\rm gas}/L)^{-1/2}$, this implies
\begin{equation}
\epsilon_{\rm ff}
\propto
\sqrt{\Sigma_{\rm gas} L}.
\end{equation}
Substituting this scaling into Eq.~\eqref{eq:sfr_master} yields
\begin{align}
\Sigma_{\rm SFR}
&\propto
(\Sigma_{\rm gas}^{1/2} L^{1/2})
\Sigma_{\rm gas}^{3/2}
L^{-1/2}
\\
&\propto
\Sigma_{\rm gas}^{2},
\end{align}
thus yielding a cloud-scale scaling consistent with Schmidt's original estimate.

\subsection{Population-averaged Scaling: Kennicutt Law ($N\approx1.5$)}

When averaging over a population of clouds within a galaxy, local variations in size, surface density, and turbulence are smoothed out. In this regime, self-regulation drives the efficiency toward a nearly constant value, and the emergent star formation law reflects the ensemble-averaged properties rather than individual cloud dynamics. If self-regulation drives the system toward a narrow range of equilibrium efficiencies across a cloud population, $\epsilon_{\rm ff}$ can be approximated as constant across the cloud population, $\epsilon_{\rm ff} \approx \mathrm{const}$. Eq.~\eqref{eq:sfr_master} then reduces to
\begin{equation}
\Sigma_{\rm SFR}
\propto
\Sigma_{\rm gas}^{3/2}.
\end{equation}
This qualitatively reproduces the observed Kennicutt relation with slope $N\approx1.4$--$1.5$. Hence, the momentum-regulated framework can reproduce the qualitative form of both the Schmidt and Kennicutt scalings under different assumptions about the variability of the star formation efficiency.

\subsection{Physical Interpretation}

The Kennicutt--Schmidt relation can be interpreted within the momentum-regulated framework as a consequence of the interplay between turbulence, gravity, and stellar feedback. On the scale of individual GMCs, the balance between feedback-driven momentum injection and turbulent dissipation sets an equilibrium velocity dispersion $\sigma_{\rm eq}$, which in turn determines the star formation efficiency. Using the virial scaling, this leads to a local relation $\Sigma_{\rm SFR} \propto \Sigma_{\rm gas}^2$, consistent with the original Schmidt law.

When averaging over a population of clouds, variations in cloud size and free-fall time partially compensate the dependence of $\epsilon_{\rm ff}$ on surface density. As a result, the effective slope is reduced, yielding a relation $\Sigma_{\rm SFR} \propto \Sigma_{\rm gas}^{1.4-1.5}$ consistent with the Kennicutt law. 

In this picture, the narrow range of observed star formation efficiencies arises from dynamical self-regulation: turbulence adjusts in response to feedback and dissipation, maintaining $\epsilon_{\rm ff}$ within a limited range despite variations in cloud properties. The framework therefore provides a simple interpretation linking cloud-scale dynamics to galaxy-scale star formation relations.

\section{Comparison with Observations}

Having established an analytic framework for momentum-regulated turbulence and its associated SFE, we now turn to a comparison between the predicted $\epsilon_{\rm ff}$ values and observational data for GMCs. This comparison aims to assess whether the simple momentum-balance picture, which incorporates stellar feedback and turbulent dissipation, is consistent with the relatively low star formation efficiencies observed in real clouds, and whether the model reproduces the weak dependence of $\epsilon_{\rm ff}$ on cloud properties. By connecting the analytic predictions to measured GMC properties such as surface density, size, and velocity dispersion, we can evaluate both the plausibility and the limitations of the momentum-regulated framework.

This comparison also allows us to assess whether additional processes, such as gravitational driving or variations in feedback strength, contribute to the observed scatter in $\epsilon_{\rm ff}$. By comparing the model with a representative sample of clouds, we can identify qualitative trends and potential deviations, providing insight into the relative importance of different mechanisms across environments. However, we emphasize that the comparison is intended as a factor-of-a-few consistency check rather than a precise fit.

\subsection{Observed Cloud Properties and Star Formation Efficiencies}

Observational surveys of giant molecular clouds (GMCs) provide a range of estimates for fundamental cloud properties, including surface density $\Sigma$, size $L$, total mass $M$, and velocity dispersion $\sigma$, as well as the star formation efficiency per free-fall time $\epsilon_{\rm ff}$ \citep[e.g.,][]{Evans2014, Heyer2015, Utomo2018, Murray2011, Kennicutt1998, Lada2010, Lee2016, Gratier2012, HuEtAl2022}. These datasets reveal that typical GMCs in the Milky Way and nearby galaxies span surface densities of roughly $50-250\ M_\odot\ \rm pc^{-2}$, radii of $10-50$ pc, masses from a few $10^3$ to $10^6\ M_\odot$, and velocity dispersions between 2 and 8 km s$^{-1}$. Correspondingly, observed SFEs per free-fall time cluster around $\epsilon_{\rm ff} \sim 0.01$, although individual clouds may exhibit scatter of a factor of a few due to environmental variations and observational uncertainties.

Table~\ref{tab:gmc_comparison} presents a selection of well-characterized molecular clouds, along with their physical properties and SFEs. The sample spans a range of surface densities ($\Sigma \sim 60$--220~M$_\odot$/pc$^2$), sizes ($L \sim 4$--52~pc), total masses ($M \sim 4\times10^3$--$6\times10^5$~M$_\odot$), and velocity dispersions ($\sigma \sim 1$--5~km/s). Observed $\epsilon_{\rm ff}$ values lie between $\sim0.5\%$ and 1.7\%, consistent with the range predicted by our model. We emphasize that the sample is intended to illustrate typical Milky Way clouds rather than to provide a statistically complete dataset. We additionally include a representative extragalactic data point based on molecular gas observations in nearby galaxies \citep{Utomo2018}, which probe cloud-scale structures on $\sim 50$--100 pc scales. Although not directly comparable to individual GMCs, this point indicates that comparable parameter regimes and SFEs may also be present outside the Milky Way. Overall, the parameter-space trends are qualitatively consistent with larger observational samples, e.g. PHANGS-ALMA \citep[e.g.,][]{Leroy2021, Sun2022, Sun2023}.

\subsection{Trends and Scatter}

Furthermore, the observational quantities listed in Table~\ref{tab:gmc_comparison} are subject to substantial uncertainties. Typical uncertainties are of order $\sim20$--50\% in cloud sizes and velocity dispersions,
$\sim0.2$--0.3 dex in surface densities, and up to $\sim0.3$--0.5 dex in the inferred SFEs. These uncertainties arise from cloud definition, projection effects, assumptions about geometry, and the indirect nature of star formation rate estimates. Consequently, the comparison presented here should be regarded as an order-of-magnitude consistency check rather than a precise quantitative fit. Deviations by factors of a few for individual clouds (e.g., Serpens) are therefore expected, given both the observational uncertainties and the simplified nature of the model.

Several qualitative trends can be identified. Higher-mass clouds tend to have larger sizes and moderately higher velocity dispersions, yet their inferred efficiencies remain within the same range as lower-mass systems. This behavior is consistent with the weak dependence of $\epsilon_{\rm ff}$ on cloud properties predicted by the model. In contrast, smaller nearby clouds such as Serpens, Auriga, and IC5146 NW exhibit lower masses and surface densities, but still fall within the same overall efficiency range.

Clouds with similar surface densities can nevertheless display modest differences in $\epsilon_{\rm ff}$ due to variations in $L$ or $\sigma$. For example, Perseus and Orion-B have comparable $\Sigma$, yet their inferred efficiencies differ slightly, reflecting differences in size and turbulence. Such variations are qualitatively consistent with the analytic scaling relations, while also highlighting the presence of intrinsic cloud-to-cloud scatter that is not explicitly captured by the model.

\begin{table*}
\centering
\begin{tabular}{lccccccc}
\hline
Reference & Cloud / Region & $\Sigma$ [M$_\odot$/pc$^2$] & $L$ [pc] & $M$ [M$_\odot$] & $\sigma$ [km s$^{-1}$] & Observed $\epsilon_{\rm ff}$ & Model $\epsilon_{\rm ff}$ \\
\hline
Murray et al. (2011) & Milky Way (GMC) & 218 & 52 & $6\times10^5$ & 3 & $0.017 \pm 0.006$ & 0.022 \\
Evans et al. (2014) & Auriga & 92 & 6 & $1\times10^4$ & 2 & $0.008 \pm 0.003$ & 0.006 \\
Evans et al. (2014) & Serpens & 135 & 4 & $4\times10^3$ & 2 & $0.014 \pm 0.005$ & 0.006 \\
Evans et al. (2014) & IC5146 NW & 59 & 5 & $5\times10^3$ & 2 & $0.005 \pm 0.002$ & 0.004 \\
Hu et al. (2022) & Perseus & 200 & 10 & $2\times10^4$ & 2 & $0.011 \pm 0.004$ & 0.010 \\
Hu et al. (2022) & Orion-B & 111 & 30 & $1\times10^5$ & 5 & $0.009 \pm 0.004$ & 0.013 \\
Hu et al. (2022) & Aquila‑S & 125 & 20 & $5\times10^4$ & 1 & $0.015 \pm 0.005$ & 0.011 \\
\hdashline
Utomo et al. (2018) & PHANGS galaxies & 80 & 100 & $6\times10^5$ & 7 & $0.007 \pm 0.003$ & 0.010 \\
\hline
\end{tabular}
\vspace{0.5em}
\caption{Comparison of observed GMC properties and star formation efficiencies with model predictions. For each cloud or region, we list the surface density $\Sigma$, size $L$, total mass $M$, and velocity dispersion $\sigma$, together with the observed star formation efficiency per free-fall time $\epsilon_{\rm ff}$ and the corresponding value predicted by our analytic model using $p_*/M_* \sim 3000\ \rm km\ s^{-1}$. The sample spans typical Milky Way GMCs as well as smaller nearby star-forming clouds, covering surface densities $\sim60$--220~M$_\odot$/pc$^2$, sizes 4--52~pc, and velocity dispersions 1--5~km/s. We additionally include a representative extragalactic data point from PHANGS observations, probing cloud-scale structures on $\sim50$--100 pc scales. The predicted efficiencies are consistent with the observed values at the order-of-a-few level. Given the substantial observational uncertainties (typically $\sim0.3$--0.5 dex in $\epsilon_{\rm ff}$), deviations at this level are not statistically significant. The comparison should therefore be interpreted as an order-of-magnitude consistency check rather than a precise quantitative fit.}
\label{tab:gmc_comparison}
\end{table*}

\subsection{Comparison with Model Predictions}

Figure~\ref{fig:epsilon_ff_dynamic} presents the time evolution of $\epsilon_{\rm ff}$ for six representative clouds with varying surface densities $\Sigma$, sizes $L$, and effective momentum injection $p_*/M_*$. The curves were obtained by numerically solving the ODE describing the momentum-regulated turbulent evolution presented in Eq.~\eqref{eq:sigma_ODE}, starting from different initial turbulent states. The shaded region highlights a typical observational range around $\epsilon_{\rm ff} \sim 0.01$. The figure shows that, for plausibly motivated GMC parameters, the momentum-regulated ODE framework is consistent with producing a limited variation of star formation efficiencies, with convergence occurring on timescales of order the free-fall time. Varying $\eta$ within the plausible range $\sim 5$--10 changes the normalization of $\epsilon_{\rm ff}$ by a factor of order unity, but does not alter the overall weak dependence on physical cloud properties or the existence of a narrow efficiency range.

Overall, the observed diversity of cloud properties and efficiencies indicates that while $\epsilon_{\rm ff}$ is typically of order $\sim 1\%$, significant cloud-to-cloud scatter is present. The momentum-regulated model is consistent with this behavior, in that it predicts a relatively weak dependence of $\epsilon_{\rm ff}$ on cloud properties and yields efficiencies of the correct order of magnitude. Consequently, the agreement suggests that our framework properly captures the leading-order physics governing star formation in GMCs.

\subsection{Dependence on Surface Density}

While Figure~\ref{fig:epsilon_ff_dynamic} illustrates the dynamical convergence toward a quasi-stable turbulent state, it does not directly show how the resulting efficiency depends on global cloud properties. This dependence is explored in Figure~\ref{fig:epsilon_ff_sigma}, which shows the predicted star formation efficiency per free-fall time, $\epsilon_{\rm ff}$, as a function of gas surface density, $\Sigma_{\rm gas}$, for a set of representative cloud models with varying sizes $L$ and momentum injection efficiencies $p_*/M_*$. The shaded orange region again indicates the observationally inferred range, while the grey band marks the typical surface density range of Milky Way GMCs ($\Sigma \sim 50$--180 $M_\odot\,\mathrm{pc}^{-2}$). The theoretical curves in the plot were obtained by solving Eq.~\eqref{eq:sigma_ODE} numerically for a different set of clouds.

The model predicts a weak increase of $\epsilon_{\rm ff}$ with surface density, reflecting the underlying scaling $\epsilon_{\rm ff} \propto \sqrt{\Sigma L}/(p_*/M_*)$. However, over the relatively limited variation of surface densities occupied by typical GMCs, this dependence remains sufficiently shallow that $\epsilon_{\rm ff}$ appears approximately constant, with values clustering around $\sim 1\%$ despite variations in cloud size and feedback strength. Outside this regime, the model predicts systematic deviations from the characteristic efficiency: at low surface densities, turbulence is weak and $\epsilon_{\rm ff}$ decreases, while at higher surface densities, the efficiency increases and may eventually exceed the observed range. This behavior is consistent with the expectation that the momentum-regulated equilibrium primarily describes moderately dense, feedback-regulated GMCs, and may not apply in more extreme environments.

We emphasize that the present comparison is intended as an order-of-magnitude consistency check. While the model captures the characteristic magnitude and weak parameter dependence of $\epsilon_{\rm ff}$, a more rigorous test would require homogeneous cloud samples and consistent measurement techniques, which is beyond the scope of this work. Nevertheless, the overall agreement demonstrates that our framework captures the leading-order behavior of star formation in typical GMCs.

\section{Discussion}

Having developed an analytic framework for momentum-regulated turbulence in GMCs and compared the predictions to observed cloud properties, we now synthesize the results, discuss their implications, and highlight the limitations and potential extensions of the model.

\subsection{Summary and Interpretation}

The ODE-based approach suggests that stellar feedback can maintain a quasi-stable turbulent state in GMCs, leading to star formation efficiencies per free-fall time of order $\epsilon_{\rm ff} \sim 0.5\% - 2\%$. Perturbation analysis around the equilibrium confirms that the solution is stable, and the associated relaxation timescale indicates that clouds return toward this equilibrium over a few free-fall times. Comparison with observed GMCs in the Milky Way and nearby galaxies shows that the predicted efficiencies lie within the range of measured values, consistent with the idea that momentum-regulated turbulence provides a simple energetic interpretation of the observed $\epsilon_{\rm ff}$ \citep[e.g.,][]{Lada2010, Murray2011, Evans2014, Lee2016, Utomo2018}.

While the framework assumes a roughly constant momentum injection per unit stellar mass ($p_*/M_*$), variations in the initial mass function (IMF) or metallicity are expected to modulate this quantity. A top-heavy IMF increases the fraction of massive stars and thus the effective feedback strength, whereas a bottom-heavy IMF has the opposite effect. Similarly, metallicity influences stellar winds and radiation pressure, altering $p_*/M_*$. Within the present framework, such variations are expected to shift the equilibrium only moderately, so that $\epsilon_{\rm ff}$ remains relatively insensitive to these parameters to first order. This behavior is consistent with observational indications across different environments \citep[e.g.,][]{Hopkins2012, Krumholz2012, Federrath2013, Krumholz2019, Kim2021}. This relative insensitivity arises because the number of supernovae per unit stellar mass varies only weakly for standard stellar IMFs, leading to comparatively small variations in the effective momentum injection $p_*/M_*$.

Our analytic approach complements previous numerical and semi-analytic studies of GMC star formation. Turbulence-regulated models \citep{Krumholz2005, Federrath2012} and simulations including detailed feedback physics \citep{Ostriker2010, Hopkins2012, Hopkins2014, Agertz2013, Grudic2018} have reproduced similarly low efficiencies. In contrast to models that rely on assumptions about the density PDF or specific collapse criteria, the present ODE framework provides a simplified dynamical picture that highlights the interplay between feedback, gravity, and turbulent dissipation. The results therefore offer a complementary perspective on how self-regulated star formation efficiencies may arise in turbulent clouds.

The framework developed in this paper can be also placed in the broader context of existing theoretical models of star formation in turbulent molecular gas. Turbulence-regulated star formation models based on the density probability distribution function (PDF), such as those developed by \citet{Krumholz2005} and \citet{Burkhart2018}, relate the star formation efficiency to the fraction of gas above a critical density threshold. In these models, $\epsilon_{\rm ff}$ depends sensitively on the Mach number and virial parameter through the shape of the density PDF. In contrast, the present framework does not explicitly model the density distribution, but instead captures the global regulation of turbulence through a balance of momentum injection and dissipation.

Self-regulation models on galactic scales, such as those by \citet{Ostriker2022} and \citet{Krumholz2018}, similarly emphasize the balance between (turbulence) feedback and gravity in setting the star formation rate. These models typically operate at kiloparsec scales and describe how feedback maintains vertical pressure balance in galactic disks. The present model can be viewed as a complementary cloud-scale analogue, in which the same physical principle (balance between feedback and gravity) is applied to the internal turbulent dynamics of GMCs.

The framework in this paper is most applicable to clouds that are moderately dense, approximately virialized, and primarily supported by feedback-driven turbulence. Extreme environments—such as very diffuse clouds, dense starburst regions, or systems with strong magnetic support or galactic shear—may lead to departures from the predicted behavior. Similarly, additional feedback channels not explicitly included in the model, such as radiative heating, ionization, or cosmic ray pressure, could modify the cloud dynamics. Observed scatter in $\epsilon_{\rm ff}$ may therefore reflect a combination of these effects, as well as intrinsic variations in cloud structure, IMF, or metallicity.

An important question concerns the interpretation of the low star formation efficiency in the context of short GMC lifetimes. Observational studies suggest that molecular clouds are transient structures with lifetimes of order $\sim 1$--3 free-fall times, and are rapidly dispersed by stellar feedback \citep[e.g.,][]{Kruijssen2019, Chevance2019, Kim2021, Chevance2022}. In this picture, the observed $\epsilon_{\rm ff} \sim 1\%$ may arise as a time-averaged property of a population of clouds, rather than reflecting a long-lived equilibrium state within individual objects.

At first glance, this interpretation appears to differ from the framework developed here, which describes the relaxation of the turbulent velocity dispersion toward a quasi-stable state set by the balance between feedback and dissipation. However, the two pictures are not mutually exclusive. In our model, the relaxation timescale of the turbulent velocity dispersion is of order the cloud crossing time, $\tau_{\rm relax} \sim L/\sigma$, which is comparable to the free-fall time under typical GMC conditions. Consequently, clouds are expected to evolve toward an equilibrium state over a significant fraction of their lifetime, even if they do not reach a strict steady state before being disrupted.

In this sense, the equilibrium solution is more appropriately interpreted as a \emph{dynamical tendency} of the system, which, when combined with short cloud lifetimes, gives rise to a \emph{statistical equilibrium} at the level of the cloud population. Individual clouds may move toward the equilibrium configuration but are dispersed before full convergence, such that the ensemble-averaged efficiency reflects the underlying dynamical regulation rather than a true steady state within each cloud.

These two interpretations may, in principle, be distinguished observationally. If individual clouds evolve toward a quasi-stable equilibrium, one expects systematic correlations between turbulent velocity dispersion, cloud properties, and instantaneous star formation activity at fixed evolutionary stage. In contrast, a purely lifecycle-driven scenario would predict larger stochastic variations and weaker correlations at fixed cloud properties, with the observed efficiency set primarily by the time distribution of cloud evolutionary phases. High-resolution observations that resolve individual clouds and trace their  evolution, as well as statistical studies of cloud populations across different environments and evolutionary stages, therefore provide a promising avenue to discriminate between dynamical self-regulation and purely population-averaged  interpretations.

\subsection{Regime of Validity and Limitations}

While the momentum-regulated framework provides a useful conceptual description of the relatively low star formation efficiencies observed in typical GMCs, it is important to delineate the range of cloud properties and environmental conditions over which the model can be expected to apply.

The analytic approach assumes that clouds are approximately virialized, with turbulence maintained primarily by stellar feedback, and that the momentum injection per unit stellar mass, $p_*/M_*$, is roughly constant. Under these assumptions, a quasi-stable equilibrium for the turbulent velocity dispersion emerges from the balance between driving and dissipation. However, significant deviations from typical GMC conditions can challenge this picture. Very low-density or diffuse molecular regions may experience turbulence decay that is not efficiently replenished by feedback, leading to departures from the predicted $\epsilon_{\rm ff}$. Conversely, extremely compact, high surface density clouds, such as those found in starburst environments, may have free-fall times shorter than characteristic feedback timescales, reducing the effectiveness of momentum-regulated self-regulation.

The model also neglects several physical processes that may become important in specific environments. Magnetic fields can provide additional support against gravitational collapse, effectively modifying the turbulent velocity scale. Large-scale galactic dynamics, such as shear or external pressure, may introduce additional driving or confinement mechanisms that are not captured by the one-zone formulation. Similarly, cosmic rays or radiative feedback processes (e.g., photoionization or heating) can contribute to the overall pressure balance, altering the effective coupling between star formation and turbulence. In the present framework, such effects would primarily enter as modifications to the effective coefficients in the dynamical equation rather than introducing qualitatively new behavior.

Observed clouds exhibit significant scatter in $\epsilon_{\rm ff}$, reflecting variations in cloud structure, evolutionary stage, and local environment. Because the momentum-regulated model predicts only a relatively weak dependence of $\epsilon_{\rm ff}$ on cloud properties, large deviations from the predicted range may indicate conditions in which the underlying assumptions are not strictly satisfied. In particular, extreme environments—such as starburst nuclei or low-metallicity dwarf galaxies—provide important regimes in which the applicability of the model can be tested.

Another important limitation of the present framework arises at very high gas surface densities, where the assumption of efficient feedback-regulated turbulence may break down. Numerical simulations have shown that above a critical surface density $\Sigma_{\rm crit} \sim 10^3$--$10^4\ M_\odot\ \mathrm{pc}^{-2}$, stellar feedback is no longer able to efficiently disrupt collapsing gas structures, leading to substantially higher integrated star formation efficiencies \citep[e.g.,][]{Grudic2018, Lancaster2021, Menon2024}. In this regime, the integrated star formation efficiency ($\epsilon_{\rm int}$) can approach values of order unity, in contrast to the low efficiencies characteristic of Milky Way GMCs. Physically, this transition reflects the fact that the free-fall time becomes shorter than the characteristic timescales for feedback to couple momentum to the gas, such that gravitational collapse proceeds more rapidly than feedback can regulate it. As a result, the momentum-regulated equilibrium described in this work ceases to apply, and the system may enter a runaway collapse regime. Thus, the breakdown of low star formation efficiencies at high surface densities does not contradict the present model, but instead marks a transition to a physical regime in which the underlying assumptions of momentum-regulated turbulence are no longer satisfied.

This distinction is particularly relevant in the context of high-redshift galaxies observed with JWST, where high gas surface densities and low metallicities may favor such feedback-limited starburst conditions \citep[e.g.,][]{Dekel2023, Somerville2025}. The present model is therefore not intended to describe these extreme environments, but rather the regime of typical, moderately dense GMCs in which feedback can effectively regulate turbulence.

Taken together, these considerations suggest that while the momentum-regulated framework captures key aspects of the physics governing star formation in typical GMCs, its predictions should be interpreted with caution outside this regime. The model is most applicable to clouds that are moderately dense, approximately virialized, and primarily supported by feedback-driven turbulence. Deviations from these conditions—whether due to extreme cloud properties, additional physical processes, or environmental effects—represent natural boundaries of the analytic framework. Identifying and characterizing these limits provides a useful context for interpreting both observations and more detailed numerical models.

\subsection{Testable Predictions}

Our framework makes a set of qualitative but testable predictions that follow directly from the momentum-regulated equilibrium.

First, the model predicts a weak but systematic increase of the star formation efficiency with gas surface density, $\epsilon_{\rm ff} \propto \sqrt{\Sigma L}$ at fixed cloud size. This scaling arises from the virial relation between turbulent velocity dispersion and surface density, implying that more strongly self-gravitating clouds reach higher equilibrium velocity dispersions and moderately enhanced efficiencies. Over the relatively narrow range of surface densities typical for Milky Way GMCs, this dependence is shallow and may appear approximately constant, but a measurable trend is expected over a broader dynamic range.

Second, in low-density or weak-feedback environments, $\epsilon_{\rm ff}$ is expected to fall below the canonical $\sim 1\%$ level. In this regime, feedback-driven turbulence is insufficient to maintain the self-regulated equilibrium, allowing turbulence to decay and reducing the effective coupling between star formation and momentum injection. As a result, clouds may evolve toward more quiescent, weakly star-forming states.

Third, the model predicts a breakdown of the momentum-regulated equilibrium at sufficiently high surface densities ($\Sigma \gtrsim 10^3$--$10^4\ M_\odot\ \mathrm{pc}^{-2}$). In this regime, the free-fall time becomes shorter than the characteristic feedback timescale, such that momentum injection can no longer efficiently regulate collapse. The system is therefore expected to transition to a regime of runaway collapse with significantly elevated star formation efficiencies, as inferred in starburst environments.

Finally, the dynamical formulation implies that the turbulent velocity dispersion relaxes toward its equilibrium value on a timescale of order the cloud crossing time, $\tau_{\rm relax} \sim L/(2\sigma)$, which is typically comparable to the free-fall time. This predicts that molecular clouds should approach their equilibrium within a few dynamical times, providing a potential observational test through measurements of cloud evolution and turbulence.

These predictions can be tested with resolved observations of molecular clouds in nearby galaxies, for example through surveys such as PHANGS-ALMA \citep[e.g.,][]{Leroy2021, Sun2022, Sun2023}, which provide measurements of gas surface densities, velocity dispersions, and star formation rates on cloud scales.

\subsection{Physical Origin of the Feedback Momentum Scale}

The characteristic momentum injection per unit stellar mass, $p_*/M_*$, plays a central role in setting the equilibrium star formation efficiency in the momentum-regulated framework. It is therefore useful to relate this quantity to the underlying physics of stellar feedback.

The dominant contribution to the momentum budget arises from supernova explosions. A single supernova releases an energy of order $E_{\rm SN} \sim 10^{51}\ \mathrm{erg}$. During the early Sedov--Taylor phase, the expansion of the remnant is approximately energy-conserving. However, as the remnant sweeps up ambient gas, radiative cooling becomes efficient and removes thermal energy from the system. At this point, the evolution transitions to a momentum-conserving phase, and the final injected momentum is effectively set. The terminal momentum can be estimated as
\begin{equation}
p_{\rm SN} \sim \sqrt{2 E_{\rm SN} M_{\rm cool}},
\end{equation}
where $M_{\rm cool}$ is the swept-up mass at the onset of efficient cooling \citep{Cioffi1988,Thornton1998}. Numerical simulations show that this leads to a characteristic momentum injection of $p_{\rm SN} \sim 2\text{--}5 \times 10^5\ M_\odot\,\mathrm{km\,s^{-1}}$, with only a weak dependence on the ambient density material, $p_{\rm SN} \propto n^{-1/7}$ \citep{KimOstriker2015,Martizzi2015}.

For a standard stellar initial mass function \citep{Kroupa2001,Chabrier2003}, the number of supernovae per unit stellar mass is approximately $N_{\rm SN}/M_* \sim 1/(100\ M_\odot)$. Combining this with the terminal momentum per supernova yields an effective momentum injection per unit stellar mass of
\begin{equation}
\frac{p_*}{M_*} \sim \frac{p_{\rm SN}}{100\,M_\odot} \sim 2000\text{--}5000\ \mathrm{km\,s^{-1}},
\end{equation}
consistent with the fiducial value adopted in this paper.

The quantity $p_*/M_*$ should be distinguished from the total feedback yield, $\Upsilon_{\rm tot}$, introduced by \citet{Ostriker2022}. While both characterize momentum injection per unit stellar mass, they represent different physical quantities. Our estimate of $p_*/M_*$ corresponds to the total momentum generated by stellar feedback, derived from the terminal momentum of supernova remnants and the stellar initial mass function. In contrast, $\Upsilon_{\rm tot}$ is an effective feedback yield calibrated from TIGRESS simulations that quantifies the momentum contributing to the turbulent support of the ISM. The factor of a few difference between the two values therefore primarily reflects their different definitions and the finite coupling efficiency of stellar feedback, rather than an inconsistency between the approaches.

This connection suggests that the characteristic magnitude of the star formation efficiency ultimately reflects the ratio between a gravitational velocity scale, $\sqrt{G \Sigma L}$, and a feedback momentum scale that is itself set by supernova physics and the stellar initial mass function. The characteristic value of $\epsilon_{\rm ff} \sim 0.01$ therefore does not arise from a fundamental constant, but from the combination of gravitational dynamics and the relatively weakly varying momentum injection associated with stellar feedback.

\subsection{Outlook and Future Directions}

While the momentum-regulated framework developed here is consistent with the observed range of star formation efficiencies in typical GMCs, several extensions remain to be explored. In particular, applying the model to more extreme environments—such as starburst regions, galaxy mergers, or high-redshift systems—may help to identify the conditions under which feedback-driven self-regulation breaks down.

Incorporating additional physical processes, including magnetic fields, cosmic rays, and time-dependent stellar feedback, could further refine the model and clarify how these mechanisms modify the balance between turbulence and collapse. Understanding how such effects influence the relaxation timescale relative to the free-fall time is a promising direction.

On the observational side, upcoming surveys with ALMA and JWST will provide improved constraints on star formation efficiencies across diverse environments. Combining resolved cloud-scale measurements with galaxy-integrated observations offers a direct test of the multi-scale connections proposed in this framework. Extending the model to link cloud-scale turbulence with global galactic dynamics represents a natural next step, with the potential to further unify local self-regulation and large-scale star formation laws.

\begin{acknowledgements}
We thank Ralf Klessen for insightful discussions and valuable feedback on the conceptual development of this project. We also thank the referee for constructive comments that significantly improved the clarity and rigor of the manuscript.
\end{acknowledgements}

\bibliographystyle{aa}
\bibliography{bibliography}

\appendix

\section{Sensitivity to Model Parameters}
\label{appendix:model_parameters}

\begin{figure*}
    \centering
    \includegraphics[width=\textwidth]{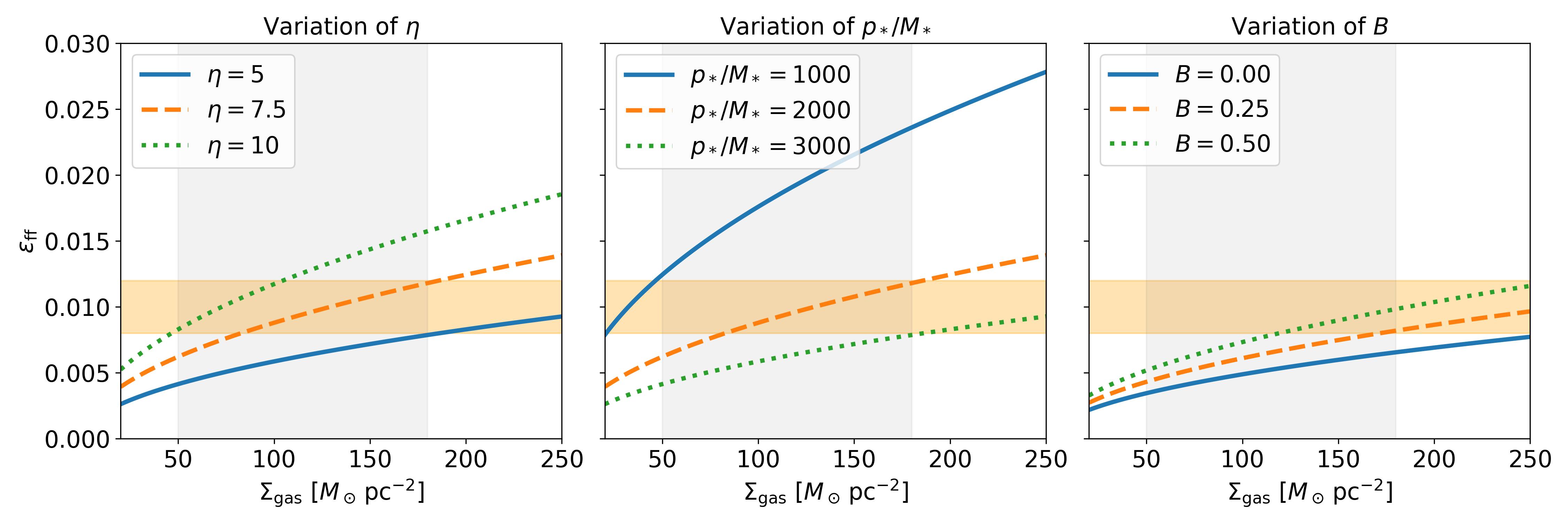}
    \caption{
    Sensitivity of the predicted star formation efficiency per free-fall time, $\epsilon_{\rm ff}$, to variations in the principal model parameters. The fiducial model adopts $L=20$~pc, $\eta=5$, $p_*/M_*=3000$~km\,s$^{-1}$, and $B=0.2$. In each panel, one parameter is varied while all others are held fixed. Left: variation of the effective feedback calibration factor $\eta$. Middle: variation of the momentum injected per unit stellar mass, $p_*/M_*$. Right: variation of the gravity-driving coefficient $B$. The shaded orange band indicates the characteristic observed range $\epsilon_{\rm ff}\sim1\%$, while the grey region marks the typical surface density range of Milky Way giant molecular clouds. Although the normalization shifts by factors of order unity, the weak dependence on surface density and the overall agreement with observations remain unchanged.
    }
    \label{fig:parameter_sensitivity}
\end{figure*}

To assess the robustness of the analytic framework, we explored the dependence of the predicted star formation efficiency on the principal free parameters of the model. Figure~\ref{fig:parameter_sensitivity} shows the resulting variation of $\epsilon_{\rm ff}$ as a function of gas surface density for a representative cloud with fiducial parameters
\[
L = 20~\mathrm{pc}, \qquad
\eta = 5, \qquad
p_*/M_* = 3000~\mathrm{km\,s^{-1}}, \qquad
B = 0.2.
\]
Each panel varies one parameter while keeping all others fixed at their fiducial values. The left panel shows the effect of changing the effective feedback calibration factor over the range $\eta = 5$--10. The middle panel varies the momentum injected per unit stellar mass over $p_*/M_* = 1000$--3000~km\,s$^{-1}$. The right panel explores the phenomenological gravity-driving coefficient over the range $B = 0$--0.5.

As expected from the analytic scaling relations, increasing $\eta$ or $B$ raises the equilibrium efficiency, while increasing $p_*/M_*$ reduces it. These parameter variations shift the predicted efficiencies by factors of order unity, but do not alter the overall weak dependence of $\epsilon_{\rm ff}$ on gas surface density. In all cases considered, the predicted efficiencies remain within the observed range of approximately $0.5$--$2\%$ for typical Milky Way giant molecular cloud conditions. This demonstrates that the principal conclusions of the model are robust to reasonable variations of the uncertain input parameters.

\section{Logarithmic Sensitivity of the Star Formation Efficiency}
\label{appendix:SFE_sensitivity}

To quantify the dependence of the star formation efficiency on the underlying model parameters, we consider the logarithmic derivatives of $\epsilon_{\rm ff}$ with respect to the relevant quantities. Starting from the scaling relation
\begin{equation}
\epsilon_{\rm ff} \propto 
\eta \,
\Sigma^{1/2}
L^{1/2}
(p_*/M_*)^{-1},
\end{equation}
we first take the natural logarithm, which transforms the multiplicative scaling into a sum of terms,
\begin{equation}
\ln \epsilon_{\rm ff}
=
\ln \eta
+
\frac{1}{2} \ln \Sigma
+
\frac{1}{2} \ln L
-
\ln (p_*/M_*).
\end{equation}
This representation makes explicit how each parameter contributes additively to the total variation of $\ln \epsilon_{\rm ff}$. Taking total differentials then yields
\begin{equation}
\frac{\delta \epsilon_{\rm ff}}{\epsilon_{\rm ff}}
=
\frac{\delta\eta}{\eta}
+
\frac{1}{2} \frac{\delta \Sigma}{\Sigma}
+
\frac{1}{2} \frac{\delta L}{L}
-
\frac{\delta (p_*/M_*)}{p_*/M_*},
\end{equation}
which directly relates fractional changes in the efficiency to fractional variations in the underlying parameters. This expression can be interpreted as a linear response relation in logarithmic space. From this, the logarithmic sensitivities follow immediately as
\begin{equation}
\frac{\partial \ln \epsilon_{\rm ff}}{\partial \ln \eta} = 1, \quad
\frac{\partial \ln \epsilon_{\rm ff}}{\partial \ln \Sigma} = \frac{1}{2}, \quad
\frac{\partial \ln \epsilon_{\rm ff}}{\partial \ln L} = \frac{1}{2}, \quad
\frac{\partial \ln \epsilon_{\rm ff}}{\partial \ln (p_*/M_*)} = -1.
\end{equation}
These coefficients quantify the relative importance of each parameter in setting the star formation efficiency. In particular, they show that $\epsilon_{\rm ff}$ depends linearly on the effective feedback calibration factor $\eta$ and inversely on the momentum injected per unit stellar mass, while the dependence on global cloud properties such as surface density and size is sub-linear. 

As a consequence, order-unity variations in $\Sigma$ or $L$ translate only into comparatively small changes in $\epsilon_{\rm ff}$, whereas variations in the effective feedback strength produce proportionally larger shifts. This provides a quantitative explanation for the weak sensitivity of the efficiency to cloud-scale properties and supports the interpretation that the observed clustering of $\epsilon_{\rm ff}$ around $\sim 1\%$ arises from the underlying scaling relations of the momentum-regulated framework.

\section{Robustness to the Choice of Closure Relation}
\label{appendix:sensitivity}

\begin{figure*}
\centering
\includegraphics[width=\textwidth]{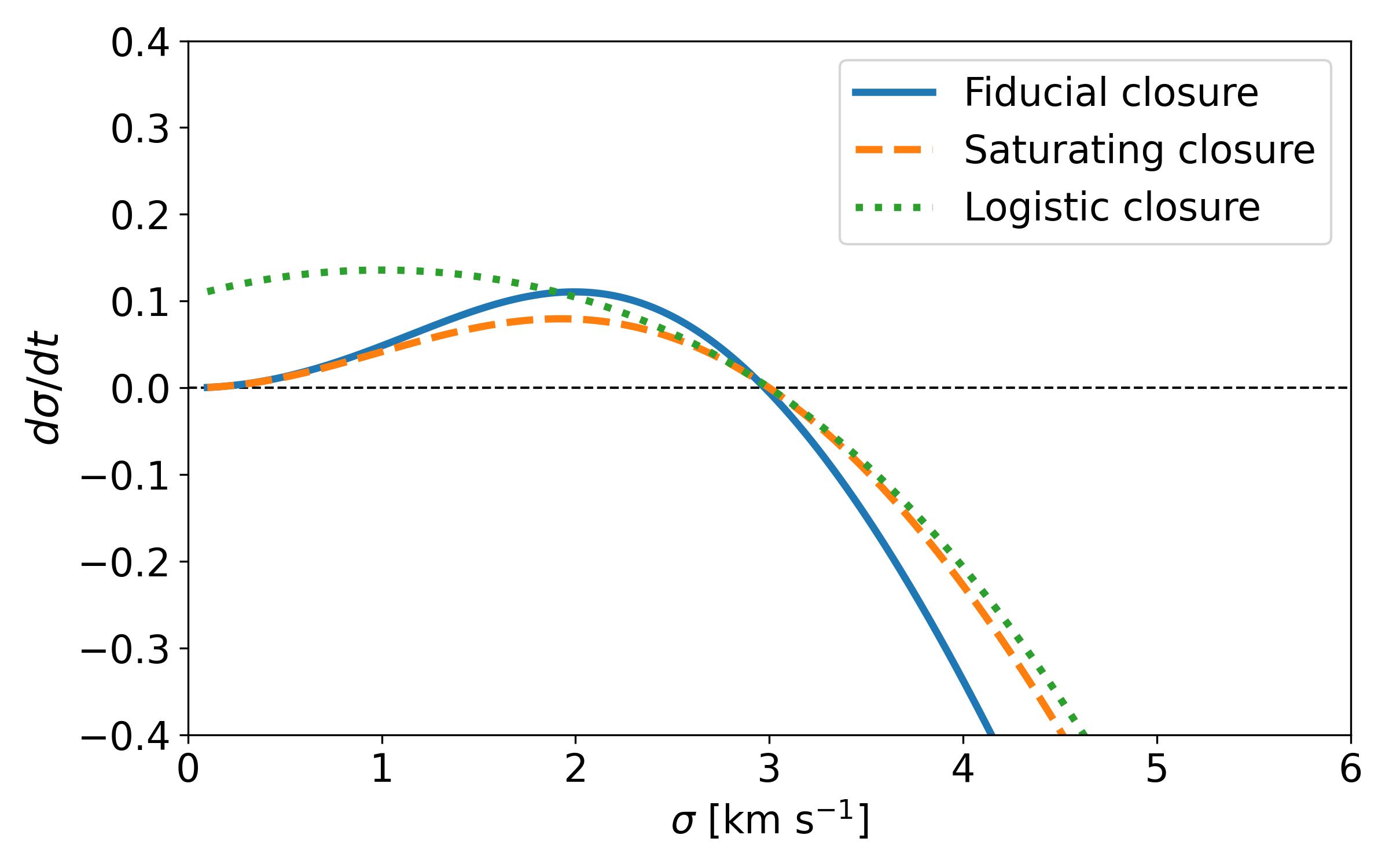}
\caption{Alternative closure relations for the star formation efficiency per free-fall time, $\epsilon_{\rm ff}(\sigma)$, used to test the robustness of the dynamical model. Shown are the fiducial closure from the main text, a saturating rational form, and a logistic turnover. All three closures produce a single stable equilibrium of the turbulent evolution equation and lead to qualitatively similar self-regulated behavior.}
\label{fig:closure_robustness}
\end{figure*}

Figure~\ref{fig:closure_robustness} shows three representative closure relations for the star formation efficiency per free-fall time, $\epsilon_{\rm ff}$, as a function of the turbulent velocity dispersion, $\sigma$. To test the robustness of the dynamical behavior discussed in the main text, we consider the fiducial closure together with two alternative phenomenological forms. The fiducial cloud parameters used for this comparison are $L = 20\,{\rm pc}$, $\Sigma = 100\,M_\odot\,{\rm pc}^{-2}$, $M = 10^5\,M_\odot$, $t_{\rm ff} = 4.4\,{\rm Myr}$, $\eta = 5$, and $p_*/M_* = 3000\,{\rm km\,s^{-1}}$, from which follow $\sigma_0 \approx 3\,{\rm km\,s^{-1}}$ as the virial velocity and $\epsilon_0 \approx 0.01$. The purpose of this appendix is not to derive a unique physical law, but to demonstrate that the existence of a stable equilibrium does not depend sensitively on the exact functional form of the closure, provided it introduces a negative coupling between strong turbulence and star formation at sufficiently large $\sigma$.

The fiducial model used in the main text is given by
\begin{equation}
\epsilon_{\rm ff}(\sigma)
=
\epsilon_0
\frac{(\sigma/\sigma_0)^2}{1+(\sigma/\sigma_0)^3},
\end{equation}
which provides a smooth turnover and suppresses star formation at large $\sigma$. As an alternative saturating closure, we consider
\begin{equation}
\epsilon_{\rm ff}(\sigma)
=
\epsilon_0
\frac{(\sigma/\sigma_0)^2}{1+(\sigma/\sigma_0)^2}.
\end{equation}
This form rises at small $\sigma$ and saturates at large $\sigma$, while preserving a single characteristic transition scale. Finally, we also test a logistic closure,
\begin{equation}
\epsilon_{\rm ff}(\sigma)
=
\frac{\epsilon_0}{1+\exp[(\sigma-\sigma_0)/\Delta\sigma]},
\end{equation}
which provides a smooth monotonic suppression of star formation at large velocity dispersion. Here, $\sigma_0$ sets the turnover scale and $\Delta\sigma$ controls the width of the transition.

For each of these closures, we integrate the dynamical equation for $\sigma(t)$ using the same fiducial cloud parameters listed above. In all cases, the system relaxes toward a single stable fixed point, with only modest differences in the equilibrium value and relaxation timescale. This demonstrates that the qualitative conclusions of the model are robust against reasonable variations in the closure prescription.

We therefore treat the fiducial closure as a representative minimal ansatz rather than as a uniquely preferred physical law. The alternative closures are included as a sensitivity test to show that the self-regulated low-efficiency state persists under plausible modifications of the functional form.

\end{document}